\documentclass[11pt,a4paper]{article}

\usepackage{jheppub}

\usepackage{amsmath}
\usepackage{verbatim}
\usepackage{amssymb}
\usepackage{inputenc}
\usepackage{textcomp}
\usepackage{appendix}
\usepackage{mathtools}

\newcommand{\e}{\epsilon}
\newcommand{\be}[1]{\begin{equation}\label{#1} }
\newcommand{\ee}{\end{equation}}
\newcommand{\bea}[1]{\begin{eqnarray}\label{#1} }
\newcommand{\eea}{\end{eqnarray}}
\newcommand{\bes}{\begin{subequations}}
\newcommand{\ees}{\end{subequations}}

\newcommand{\p}{\partial}
\newcommand{\refb}[1]{(\ref{#1})}

\renewcommand{\O}{{\mathcal{O}}}
\renewcommand{\L}{{\mathcal{L}}}

\newcommand{\bL}{\bar{{\mathcal{L}}}}

\renewcommand{\>}{\rangle}

\newcommand{\non}{\nonumber}

\renewcommand{\)}{\right)}

\renewcommand{\a}{\alpha}

\newcommand{\lb}{\left[}
\newcommand{\rb}{\right]}

\title{Galilean Yang-Mills Theory}

\author[a]{Arjun Bagchi,} \author[b]{Rudranil Basu,} \author[c]{Ashish Kakkar,} \author[c, d]{and Aditya Mehra}

\affiliation[a]{Center for Theoretical Physics, Massachusetts Institute of Technology,\\ 77 Massachusetts Avenue, Cambridge, MA 02139, USA.\\} 

\affiliation[b]{Saha Institute of Nuclear Physics\\ Block AF, Sector 1, Bidhannagar, Kolkata 700068. INDIA. \\}

\affiliation[c]{Indian Institute of Science Education and Research\\ Dr Homi Bhabha Road, Pashan. Pune 411008. INDIA.\\} 

\affiliation[d]{Van Swinderen Institute for Particle Physics and Gravity, University of Groningen, \\ Nijenborgh 4, 9747 AG Groningen, The Netherlands.\\ }

\emailAdd{abagchi@mit.edu, rudranil.basu@saha.ac.in, ashishk@students.iiserpune.ac.in,\\ aditya.mehra@students.iiserpune.ac.in}

\abstract{We investigate the symmetry structure of the non-relativistic limit of Yang-Mills theories. Generalising previous results in the Galilean limit of electrodynamics, we discover that for Yang-Mills theories there are a variety of limits inside the Galilean regime. We first explicitly work with the $SU(2)$ theory and then generalise to $SU(N)$ for all $N$, systematising our notation and analysis. We discover that the whole family of limits lead to different sectors of Galilean Yang-Mills theories and the equations of motion in each sector exhibit hitherto undiscovered {\em{infinite dimensional symmetries}}, viz. infinite Galilean Conformal symmetries in $D=4$. These provide the first examples of {\em{interacting}} Galilean Conformal Field Theories (GCFTs) in $D>2$. }

\preprint{MIT-CTP/4755}

\begin{document}
\maketitle

\section{Introduction}
Yang-Mills (YM) theories form the backbone of our understanding of the present observed universe. Three of the four fundamental forces, the electromagnetic, the weak and the strong forces, are described by quantum YM theories. Even the forever truant gravity, whose union with quantum mechanics has been an unhappy one, can be understood in terms of YM theories in the new light of the Holographic Principle \cite{{'tHooft:1993gx, Susskind:1994vu}}. The Holographic Principle relates a theory of quantum gravity in a certain dimensional space-time to a theory without gravity living on the boundary of this space-time. Its most well-understood avatar, the AdS/CFT correspondence \cite{Maldacena:1997re}, in its most familiar setting, is a mapping between a string theory (Type IIB) living on five dimensional Anti de Sitter (AdS) space-times (times a five sphere ($S^5$)) and $\mathcal{N}=4$ SU(N) Supersymmetric Yang-Mills (SYM) theory which is a 4 dimensional (4d) conformal field theory living on the boundary of AdS$_5$. So understanding YM theories can teach us about quantum gravity in this very unique and non-intuitive way. 

Hence, it is obvious that the study of YM theories is central to the understanding of the workings of nature and over the decades since its discovery in the 1950s, there is a very large body of work which addresses different aspects of classical and quantum YM theory. Often it is very useful to look at effective theories which are descriptions of the full theory restricted to a certain regime in parameter space. The Fermi theory for electro-weak interactions is a good example of this. Although this was not the correct theory which explained the W and Z bosons, the theory is very good as an effective field theory for energies well below the formation of these bosons.  

It is also at times very illuminating to look at limits of fundamental theories to discover perhaps a closed sub sector where the theory becomes more tractable. A prime example of this is the planar limit ($N \to \infty$ with the rescaled coupling constant $\lambda = g_{YM}^2 N$ held fixed) of $SU(N)$ $\mathcal{N}=4$ SYM, which leads to an integrable sub-sector of the theory \cite{Beisert:2010jr}. In our present work, we shall be interested in such a limit of classical YM theory, a non-relativistic limit where we send the speed of light to infinity that can be looked upon as an effective theory when the degrees of freedom of interest move at very low speeds. 

\subsection*{Non-relativistic limit of Yang-Mills theory}
As just advertised, in this paper we would be interested in constructing the non-relativistic limit of Yang Mills theory. This is a generalisation of our earlier work on the construction of the systematic non-relativistic limit of Electrodynamics in \cite{Bagchi:2014ysa} (following earlier work \cite{LBLL}) and the motivations remain the same. 

Classical Yang Mills theories exhibit conformal invariance in $D=4$. It is thus expected that the Galilean version of YM theory will exhibit a similar non-relativistic conformal symmetry in $D=4$. This symmetry is governed by the so-called Galilean Conformal Algebra (GCA), the finite part of which arises from a contraction of the usual relativistic $d$ dimensional conformal symmetry $SO(D,2)$. However, there is more to the non-relativistic symmetry than just the contraction. It was observed in \cite{Bagchi:2009my} that the GCA can actually be given an infinite dimensional lift in any space-time dimension. This enhancement of symmetry in the non-relativistic theory was indeed one of the very novel claims of \cite{Bagchi:2009my}. The claim obviously needed to be justified by explicit examples of physical systems where this infinite symmetry is realised. It was mentioned in \cite{Bagchi:2009my} that the Euler and Navier-Stokes equations in non-relativistic hydrodynamics exhibit symmetries under an infinite subalgebra of these symmetries which are physically represented by time dependent boosts. All quantum field theories have a hydrodynamic regime. So it seems that we have a partial realisation of the infinite GCA in the non-relativistic limit of all quantum field theories in the hydrodynamic regime. But the drawback is that this is only a partial realisation.  

In two dimensions, the GCA turns out to be a contraction of linear combinations of the two copies of the Virasoro algebra \cite{Bagchi:2009pe}. Interestingly, the asymptotic symmetry algebra of three dimensional Minkowski spacetime at null infinity, the BMS$_3$ algebra \cite{Bondi:1962px, Sachs:1962zza, Barnich:2006av} is isomorphic to the 2d GCA \cite{Bagchi:2010eg}. The 2d GCA is thus central to the understanding of holography of 3d flat spacetimes and forms the symmetry algebra of putative dual 2d field theories that live on the boundary of 3d flat space. This has been used in several recent works on the subject, a collection of which are \cite{Bagchi:2012yk} -- \cite{Barnich:2012rz}. For a more comprehensive and up to date list of references on flat holography, the reader is referred to \cite{Basu:2015evh}. In \cite{Barnich:2012rz}, examples of such field theories were constructed as limits of Liouville theory. These are thus explicit examples of 2d Galilean Conformal Field Theories (GCFTs){\footnote{The 2d GCA has also shown up as the residual symmetries on the world-sheet of the tensionless closed bosonic string \cite{Bagchi:2013bga, Bagchi:2015nca}.}}. Although the field theories discussed in \cite{Barnich:2012rz} have infinite dimensional symmetries, given that these are 2d theories arising as limits of relativistic CFTs and relativistic conformal invariant theories in 2d always have infinite symmetry, the extended symmetry structure is not a surprise.    

The infinite enhancement of symmetries in the non-relativistic limit is a non-trivial statement in $D>2$ and until recently the search for field theories exhibiting such symmetry structures was elusive. However, in \cite{Bagchi:2014ysa}, we found that the entire infinite dimensional symmetry is realised in the non-relativistic version of Maxwell's equations and this became the first known example of a GCFT in dimensions higher than two. The discovery of this new infinite-dimensional symmetry of electrodynamics opened up interesting possibilities. One of the tantalising aspects of this project is the prospect of the discovery of some new integrable sector in the theory in this non-relativistic limit. On the other hand, one of the possible criticisms of the construction in the case of source-less electrodynamics is that the system is non-interacting and perhaps the enhancement of the symmetries has something to do with the fact that it is essentially a free field theory. Skeptics would then claim that these extra symmetries would disappear the moment interactions are turned on. It could be that this symmetry has something to do with the infinite dimensional higher spin symmetry that free systems at times exhibit. We put forward some robust evidence against this in \cite{Bagchi:2014ysa}, but a concrete demonstration would constitute the construction of an explicit interacting example.  

The natural set-up to address this question of whether the above described infinite dimensional non-relativistic conformal  symmetries only arise in free theories is thus to move to YM theories where even without matter fields there are interactions due to the gauge fields at the classical level. This is the aim of  of the present paper. Generalising our construction of Electromagnetism in \cite{Bagchi:2014ysa}, we discuss how one would systematically implement the non-relativistic limit in the Yang Mills theories. 

In the case of Electrodynamics, we found in \cite{Bagchi:2014ysa}, in keeping with old literature \cite{LBLL}, that there were two distinct non-relativistic limits that one could take,- {\it{viz.}} the electric and the magnetic limits. In the YM case, together with these ``vanilla" limits, in the present analysis we find several skewed limits depending on the scaling of the different components of the gauge fields. So we are led to different sectors of the Galilean YM theory. We construct the equations of motion for all these sectors and find that in $D=4$, there are finite enhancements which are the Galilean analogues of the relativistic conformal invariance. Surprisingly, we also find that the infinite enhancements survive when one generalises electrodynamics to YM theories. 

The infinite dimensional symmetries we discussed in \cite{Bagchi:2014ysa} and the ones we discuss in this paper are classical symmetries which we expect to become anomalous in the quantum regime. But given that the infinite symmetries survive in a theory with interactions, there is the very real hope of finding these symmetries even in the supersymmetric versions of YM theory. In $\mathcal{N}=4$ SYM, relativistic conformal invariance exists in the full quantum theory. Hence the hope is that when we look at the Galilean version of $\mathcal{N}=4$ SYM, we would find similar infinite dimensional enhancements of the (super-) GCA at the fully quantum level. Aspects of the supersymmetrisation of the GCA has been dealt with in \cite{Bagchi:2009ke, Sakaguchi:2009de, deAzcarraga:2009ch, Mandal:2010gx}. One of the principal goals of our programme is to uncover quantum infinite dimensional Galilean conformal symmetries in SYM. But we will leave investigations of the SUSY version to future work and continue to build with our explorations of the bosonic case at present. 

\subsection*{Plan of the paper}
The present paper is structured in the following way. We start in Sec 2 with a recapitulation of our earlier work on the Galilean Electrodynamics to set the stage and notation for the rest of the paper. Here we cover the basics of the algebra and its representation theory focusing on the scale-spin highest weight representations introduced in \cite{Bagchi:2014ysa}. We then discuss the Galilean limit of Electrodynamics and investigate the symmetries of the equations of motion. We also address the question of gauge invariance which was not dealt with in \cite{Bagchi:2014ysa}. We show how to obtain Galilean gauge invariance as a limit of the relativistic gauge invariance and then also obtain the same results by performing an intrinsic analysis. Finally, we show how to address the question of conformal invariance of relativistic Yang Mills theories by considering equations of motion. 

In Sec 3, we construct the $SU(2)$ Galilean theory in detail. There are four distinct limits in the Galilean sector which we cover one after the other. In each sub-sector, we study the equations of motion, aspects of gauge invariance (here we use only the limiting scheme) and then analyse the symmetries of the equations of motion, first the finite and then the infinite dimensional, in detail. 

In Sec 4, we generalize our construction to the SU(N) case systematizing our explicit construction of the previous section. The general structure helps us shed light on some issues which were apparently surprising in the SU(2) case. We follow the same programme,- first detailing the scaling, then looking at the equations of motion, addressing gauge invariance before finally exhibiting the infinite dimensional symmetry of the equations of motion.  

We conclude in Sec 5 with a summary of our results, discussions and a list of possible future directions. Appendix A is a description of the action of the negative modes of the symmetry algebra on the equations discussed in the main text.

\newpage

\section{Setting the stage}
In this section we discuss and revisit the essential ingredients for our analysis of Galilean Yang Mills theories, viz. we start with a description of the required representation theory of the GCA, review the non-relativistic limit of Maxellian Electrodynamics and the emergence of the infinite Galilean conformal invariance. We then provide a quick summary of the relativistic conformal symmetry that arises in Yang-Mills theories in $D=4$. All of these would be used in the coming sections when we construct the Galilean Yang-Mills theories.  

\subsection{The infinite Galilean conformal symmetry}
The group of conformal transformations of the $D$ dimensional Minkowski space $ \mathbb{R}^{D-1,1}$ is $SO(D,2)$. The most obvious way the group of Galilean transformations is obtained is by a Inonu-Wigner contraction of this group. A more physical space-time interpretation of this procedure can be gained by noticing that the generators of the original conformal group can be represented as vector fields $ f^{\mu} (x) \p _{\mu}$ on $ \mathbb{R}^{D-1,1}$. As is evident, the process of going to the Galilean framework involves breaking of explicit Lorentz covariance in the following space-time contraction:
\be{sptmcontract}
x^i \rightarrow \epsilon x^i, \, \, t \rightarrow t , \, \, \epsilon \rightarrow 0.
\ee
This scaling of spatial coordinates means including only slow observers as $ v^{i} \sim \frac{x^i}{t} \rightarrow \e v^i$ in units of speed of light ($c=1$), thus invoking the principle of Galilean relativity. Let's describe how the space-time contraction works for vector fields generating transformations through an example of the boost generator. The Lorentz boost generator changes as
$$B_i = t \p _i + x_i \p _t \mapsto \epsilon ^{-1} t \p _i + \epsilon x_i \p _t$$ under the scaling \eqref{sptmcontract}. In order to extract the finite part of it, we define the Galilean boost multiplying this by $ \epsilon$ and the taking the appropriate limit. This results in $B_i = t \p_i$. This algorithm of `Galileanization' can be carried out for all the generators (Poincare, dilatation and special conformal). As is evident from the example of boost generator, the vector field form of the generators modify and hence do their Lie brackets, resulting a new Lie algebra, different from $\mathfrak{so}(D,2)$, which we name as finite Galilean conformal algebra (f-GCA). A basis for this algebra is spanned by the vector fields:
\be{suggen}
L^{(n)} = -t^{n+1} \p_t - (n+1) t^n x_i \p_i  \quad M^{(n)}_i = t^{n+1} \p_i ~~ \mbox{for}~~ n=0, \pm 1 \quad \mbox{ and } J_{ij} = x_{[i} \p_{j]}. 
\ee
A more familiar identification is $L^{(-1,0,1)}=H,D,K$ and $M^{(-1,0,1)}_i =P_{i},B_{i},K_{i}$ where $H, D$ and $K$ are respectively the Galilean Hamiltonian, dilatation and ($SO(D-1)$-scalar) special conformal transformation. On the other hand $P_i, B_i$ and $K_i$ represent momentum, Galilean boost and ($SO(D-1)$-vector) special conformal transformation. $J_{ij}$, as usual, generates homogeneous $SO(D-1)$ rotations. 

Working out the Lie-brackets of the vector fields \eqref{suggen} we can write the full algebra of f-GCA as      
\bea{GCA}
&& [L^{(n)}, L^{(m)}] = (n-m)  \, L^{(n+m)}, \quad [L^{(n)}, M^{(m)}_i] = (n-m) \, \, M_i^{(n+m)}, \quad [M^{(n)}_i, M^{(m)}_j]=0 \non \\
&& [J_{ij}, J_{kl}] = \delta_{k [i} J_{j] l}-\delta_{l [i} J_{j] k} , \quad [L^{(n)}, J_{ij}] = 0, \quad  [M^{(n)}_i, J_{jk}] = M_{[k}^{(n)} \delta^{}_{j]i}.
\eea
with $n,m=0,\pm 1$. One very interesting observation of \cite{Bagchi:2009ca} is that the algebra \eqref{GCA} closes even if we let the index $n$ of \eqref{suggen} run over all integers. This infinitely enhanced Lie algebra will be referred to as GCA from now on{\footnote{An even larger infinite algebra can be obtained if we give a lift to the rotation generators $$J_{ij}^{(n)}= t^n x_{[i} \p_{j]}.$$ We shall however choose not to work with this larger algebra as it does not turn out to leave theories under consideration invariant. As of now, we don't understand the reason behind this.}. The embedding of f-GCA inside GCA is therefore similar to that of $SL(2, \mathbb{R})$ in Witt algebra (algebra of smooth diffeomorphisms of $S^1$). Thus there seems to be an infinite enhancement of symmetries when one looks at non-relativistic limits of conformal field theories in any dimension. This claim obviously needs to be justified by looking at examples. A partial realisation was achieved in \cite{Bagchi:2009my} when one considered non-relativistic hydrodynamics. Here one found that the Euler and the Navier-Stokes equations were invariant under arbitrary time dependent boosts, or in the language of the algebra, invariant under all $M^{(n)}_i$. But until recently, there was no example of a theory which realised the full GCA as its symmetry in dimensions $D>2$. Galilean Electrodynamics, as we go on to describe now, was the first example of a GCFT in $D>2$. 

\subsection{Scale-spin highest weight representation of GCA}
To set the stage, we need to discuss aspects of the representation theory constructed in \cite{Bagchi:2014ysa}. We will be interested in the scale-spin highest weight representations, where the states are labelled by weights under the dilatation and rotation generators as opposed to the scale-boost representations of \cite{Bagchi:2009ca} which, e.g. are of fundamental relevance in the $D=2$ case \cite{Bagchi:2009pe}. For further discussion on this, the reader is referred to \cite{Bagchi:2014ysa}. As just stated, we would label our states by the weights under $L^{(0)}$ and $J_{ij}$: 
\be{}
L^{(0)} | \Phi \> = \Delta  | \Phi \>, \quad J_{ij}  | \Phi \> = \Sigma _{ij}  | \Phi \> .
\ee
Then we define the primary states in a way similar to usual conformal field theories by demanding that the spectrum be bounded from below and hence these primary states are annihilated by the annihilation operators of the algebra. The primary state conditions are then given by: 
\be{posmode}
L^{(n)} | \Phi \>_p = M_i^{(n)} | \Phi \>_p = 0 \quad \forall n >0 .
\ee
To study the action of the GCA on the operators, we propose a state-operator correspondence, again in close analogy with conformal field theories, in order to have a relation between primary state and the vacuum:
\be{}
| \Phi \>_p = \Phi (0,0) | 0 \> .
\nonumber
\ee
The action of f-GCA on primaries is given by: 
\bes 
\bea{}
&&\left[J_{ij}, \Phi(0,0)\right] =  \Sigma _{ij} \Phi(0,0), \quad \left[L^{(0)}, \Phi(0,0)\right] = \Delta \Phi(0,0), \\
&&\left[L^{(-1)}, \Phi(t,x^i)\right] = \partial _t \Phi(t,x^i), \quad \left[M^{(-1)}_i, \Phi(t,x^i)\right]= -\partial _i \Phi(t,x^i), \label{mom}\\
&&\left[L^{(+1)}, \Phi(0,0)\right] = 0 = \left[M^{(+1)}_i, \Phi(0,0)\right] .
\eea
\ees
At general space-time points $(t,x^i)$, it is straightforward to work out the action of the generators of the GCA on operators:   
\be{}
\Phi(t,x) = U \Phi(0,0) U^{-1} \hspace{.2cm}\text{with}  \hspace{.4cm } U= e^{(t L^{(-1)} - x^i  M^{(-1)}_i)}
\ee
For a general GCA element $ \O$, we have
\be{}
\left[ \O, \Phi(t,x^i)\right] = U \left[ U^{-1} \O U , \Phi(0,0)\right] U^{-1} \nonumber
\ee
and then we shall use the Baker-Campbell-Hausdorff formula (BCH) and GCA \eqref{GCA} to evaluate $U^{-1} \O U$.
We looked into the subtleties regarding the action of boost  on operators in \cite{Bagchi:2014ysa} and found that when we restrict ourselves to the cases when the primaries are spin 0 and spin 1, the scale-spin representations are labelled by two constants $(r,s)$ by the action of the boost generators{\footnote{$(r,s)$ were called $(a,b)$ in \cite{Bagchi:2014ysa}. Here we rename them to avoid conflict of notion with the gauge index to be used throughout the paper.}}. The representation theory does not fix these numbers and they have to be determined by the inputs from dynamics.

Following from above, the infinite extension of GCA when acted on the operators at general space-time points would give:      
\bes\label{inf}
\bea{}
\lb L^{(n)}, \phi(t,x^i) \rb &=& t^n \(t \p _t + (n+1) x^i\, \p_i + (n+1)\Delta \) \phi - s \,n\,(n+1)t^{n-1}\,x^i \phi _i \\
\lb L^{(n)}, \phi_i(t,x^i) \rb &=& t^n \(t \p _t + (n+1) x^j\, \p_j + (n+1)\Delta \)\phi_i - r\,n\,(n+1)t^{n-1}\,x_i \phi \\
\lb M^{(n)}_i, \phi(t,x^i) \rb &=& -t^{n+1} \p_i \phi + s \,(n+1)\,t^{n} \phi_i \\
\lb M^{(n)}_i, \phi_j(t,x^i) \rb &=& -t^{n+1} \p_i \phi_j + r\,(n+1)\,t^{n} \delta_{ij}\phi
\eea
\ees
In the above, we have suppressed the spacetime dependance of the fields on the right hand side of the equations. Thus the scale-spin representations of the GCA are defined by the set $\{r,s,\Delta,\Sigma\}$.  
 
 \smallskip

\subsection{Galilean conformal invariance of Galilean Electrodynamics} \label{inveom}
We have already seen that in the non-relativistic limit of a conformal field theory, there is the conjectural infinite enhancement of symmetries in any dimensions and that this is partially realised in non-relativistic hydrodynamics. It is obviously very important to construct dynamical systems exhibiting the full symmetry, if we are to lend credibility to this claim of infinite enhancement of symmetries. One of the obvious candidates to construct such a theory is 4d Maxwellian Electrodynamics. Electrodynamics, the theory of free spin-1 bosons with $U(1)$ gauge invariance has conformal symmetry in 4 dimensional space-time at the classical level. We thus expect that the non-relativistic version of Maxwell's theory to exhibit non-relativistic conformal invariance in $D=4$. The principle question is to check whether the conjectural infinite symmetries are realised here. Interestingly, there exist two well-understood Galilean limits of the theory \cite{LBLL, Duval:2014uoa}. Let us briefly describe them, in a $D$ dimensional space-time.

The present description assumes the existence of the potential formulation of Electrodynamics. As is apparent from different scaling \eqref{sptmcontract} of space and time, when starting from a relativistic theory, the first thing to consider in a Galilean system is the breaking of Lorentz covariance. In this sense there are two possible ways of bringing in different scaling rules of the original 4-vector potential $A_{\mu}$:
\bes \label{e-m_scaling}
\bea{}
\hspace{-2cm}\mbox{{\bf{Electric limit}}:}&&\quad A_t \rightarrow A_t, \, \, A_i \rightarrow \epsilon A_i  \label{e-scaling} \\
\mbox{{\bf{Magnetic limit}}:}&& \quad A_t \rightarrow \epsilon A_t, \,  \, A_i \rightarrow  A_i  \quad \mbox{with } \, \, \epsilon \rightarrow 0. \label{m-scaling} 
\eea 
\ees
It was described in \cite{Bagchi:2014ysa} that $A_0$ and $A_i$ transform as true scalar and vector under $SO(D-1)$. From a purely representation theory point of view, the above two limits \eqref{e-m_scaling} correspond to two inequivalent representations of the Galilean boost on the space of $SO(D-1)$ tensors.

Another way of looking at the scenario is the following \cite{LBLL}. The first scaling corresponds to an extreme time-like and the second one an extreme space-like vector from a Lorentzian point of view. Electric and Magnetic fields constructed out of these scalar and vector potentials in the two scaling limits behave respectively as $ |\mathbf{E}| \gg | \mathbf{B}|$ and $ |\mathbf{E}| \ll | \mathbf{B}|$. Therefore, these two disconnected branches of Galilean electrodynamics are the Electric and the Magnetic limits \cite{LBLL}. We shall call them the Electric and Magnetic sectors in our discussions in this paper. 

The equations of motion of Galilean Electrodynamics (in absence of sources) in the two limits are respectively:
\bes
\bea{}
\hspace{-2cm}\mbox{{\bf{Electric sector}}:}\qquad &&\partial^i \partial_i A_t = 0, \quad \partial ^j \partial_j A_i - \partial _i \partial_j A^j + \partial_t \partial _i A_t= 0; \label{Eeom} \\
\mbox{{\bf{Magnetic sector}}:} \qquad &&(\partial ^j \partial_j)A_{i}-\partial_{i}\partial_{j}A^{j} =0, \quad (\partial^i \partial_i)A_{t}-\partial_{i}\partial_{t}A^{i}= 0. \label{Meom}
\eea \ees
We now wish to understand how to check that these equations are invariant under the whole GCA. Let us stress that we would be checking the invariance of the theory under the symmetries by looking at the invariance of the equations of motion{\footnote{Although it is perhaps desirable to check for symmetries at the level of the action, the non-relativistic limit makes writing an action difficult as the metric becomes degenerate. It is possible that by looking at a suitable reformulation, possibly through Newton-Cartan structures, one would be able to re-derive our results in an action formulation. This is something we would look to clarify in subsequent work.}}. In order to check symmetries of equations, it is important to lay down the rules of the game. We need to check whether the equations continue to hold with transformed field variables $\Phi(t,x)$. If an equation of motion has the schematic form: 
$ \square \Phi(t,x) =J$ then if the following also holds: 
\be{rule}
\square \delta_{ \O} \Phi(t,x) = \square \lb \O , \Phi (t,x) \rb =0, 
\ee
we would have shown that the equation has the proposed symmetry. $\O$ here denotes any relevant transformation generator. We would be using \refb{inf} for the expression of $\lb \O , \Phi (t,x) \rb$. We note that any source $J$ in the right hand side should anyway be annihilated by transformation operators, since they are non-dynamical. 

To check for invariance of Galilean Electrodynamics in the two aforementioned limits, we treat $A_0$ and $A_i$ as scalar and vector primaries. As mentioned before, we need the set $\{r,s,\Delta,\Sigma\}$ to specify the representation theory. These are obtained by looking carefully at the contraction of the relativistic theory and it turns out that 
\bea{}
\hspace{-3cm}\mbox{{\bf{Electric sector}}:}\ && \{r_e,s_e,\Delta(A_t), \Delta(A_i) \} = \left\{-1, 0, \frac{D-2}{2}, \frac{D-2}{2} \right\}. \\
\mbox{{\bf{Magnetic sector}}:} \ && \{r_m,s_m,\Delta(A_t), \Delta(A_i) \} =  \left\{0, -1, \frac{D-2}{2}, \frac{D-2}{2} \right\}.
\eea
It is now straight-forward to check for the invariance of the equations \refb{Eeom} and \refb{Meom} under the infinite dimensional symmetries using \refb{rule} and \refb{inf}. We leave it to the readers to check this or look at \cite{Bagchi:2014ysa}.  A point of interest is that the invariances hold for only $D=4$, something that was to be expected from the relativistic theory.   

\subsection{The issue of gauge invariance}\label{gauge}
One subject that was not addressed in any detail in \cite{Bagchi:2014ysa} was the issue of gauge invariance. Here we make an effort to clarify some aspects of this. In the Abelian case, the gauge transformations have the form  
\be{relga} A_\mu (x) \rightarrow A'_\mu(x) = A_\mu(x)+ \partial_\mu \alpha(x)\ee 
with $\a(x)$ being an arbitrary function. This leaves the electromagnetic action as well as the equations of motion invariant. We now would try to make sense of a non-relativistic version of gauge transformations for both the Electric and Magnetic sectors. At first, we attempt to understand it from the point of view of a limit of the relativistic theory and then will try an intrinsic Galilean analysis. 

\paragraph{{Galilean gauge transformations as a limit:}} We begin our discussions with the Electric sector where the gauge fields scale according to \refb{e-scaling}. Together with this we have the usual non-relativistic scaling of the spacetime \refb{sptmcontract}. We insert these scalings into the equation \refb{relga} and demand that this be non-singular. This fixes the scaling of the gauge parameter $\a(x)$. In the Electric sector, the gauge parameter scales as
\be{}
\a_e (t, x^i) \to \e^2 \a_e(t, x^i)
\ee
The gauge transformation in the electric limit thus takes the form
\be{elga} A_t (t, x^i) \rightarrow A_t(t, x^i), \quad A_i (t, x^i) \rightarrow A_i(t, x^i)+ \partial_i \alpha_e(t, x^i). \ee
It can easily be checked that the Electric sector equations of motion \refb{Eeom} are invariant under this set of gauge transformations. 

A similar analysis for the Magnetic sector yields the scaling for the gauge parameter
\be{}
\a_m (t, x^i) \to \e \ \a_m(t, x^i)
\ee
The gauge invariance in this limit is different and reads
\be{maga} 
A_t (t, x^i) \rightarrow A_t(t, x^i)+ \partial_t \alpha_m(t, x^i), \quad A_i (t, x^i) \rightarrow A_i(t, x^i)+ \partial_i \alpha_m(t, x^i). 
\ee
Again, it can be readily checked that these modified gauge invariances leave the magnetic equations of motion \refb{Meom} invariant. 

\paragraph{{Galilean gauge transformations as an intrinsic property:}} When we deal with intrinsically non-relativistic theories which are built on the symmetries of the Galilean group, we could ask if there are gauge symmetries in the Galilean theory, viz. transformations that alter the intrinsic variable but leave observables and equations of motion invariant. 
  
To pinpoint the forms of the gauge transformations we will resort to the following guiding principles. They
\begin{enumerate}{}
\item \label{cond1} should keep the equations of motion invariant.
\item \label{cond2} should not talk with global space-time transformations, i.e. their action on field space (strongly) should commute with at least the global part of the GCA generators. This may be relaxed when we consider gauging the global space-time transformation to include gravity and supersymmetry.
\item \label{cond3} Should be field independent. This also gets relaxed in some cases, when one enlarges space of gauge transformation, for example, including BRST.
\end{enumerate}

Let us consider Galilean electrodynamics for the moment. A general set gauge transformations for $A_0, A_i$ may be written as:
\bes \label{gt}
\bea{} 
A_t &\rightarrow & A_t + \partial_t \Lambda_1(t, x^i) \\
A_i &\rightarrow & A_i + \partial_i \Lambda_2(t, x^i)
\eea
\ees
These are chosen to be manifestly state independent, as the theory we are considering is linear. Note that due to absence of Lorentz invariance in the Galilean theory, we have this freedom of introducing two independent gauge parameters $\Lambda_1$ and $\Lambda_2$ differently. Although this is true, this freedom should be restricted by condition \ref{cond2} above. This is because there is still Galilean boost which partially mixes the scalars and vectors. Additional possible sources of restriction are from the obvious demand that \refb{gt} should keep the equations of motion invariant according to condition \ref{cond1}. Let's try to see in a step-by-step manner, what this restriction implies in the Galilean context.

Start with the well-understood relativistic case. We will be guided by the basic principle that gauge transformations don't talk with space-time transformations. To implement this analytically, let $\delta_{\omega}$ denote the Lorentz transformation by a parameter $\omega ^{\mu \nu}$:
\be{} 
\delta_{\omega}A_{\mu} = \omega^{\rho \nu} \left[x_{[\rho} \partial_{\nu]}A_{\mu} + \eta_{\mu [ \rho} A_{\nu]}\right]
\ee 
and $\delta_{\Lambda}$ be gauge transformation: $ \delta_{\Lambda}A_{\mu} = A_{\mu}+\partial_{\mu} \Lambda$. The condition of the independence of the gauge and space-time transformations would therefore hold if they commute: 
\be{}
(\delta_{\omega} \delta_{\Lambda} - \delta_{\omega} \delta_{\Lambda})A_{\mu}=0.
\ee
It can be easily checked, but has a small subtlety in the evaluation of $\delta_{\omega}\Lambda$. Although $\Lambda$ is a parameter, it is dynamical since we have not gauge fixed the system and behaves like a scalar under $\delta_{\omega}$. With this consideration, we see indeed that the above commutation holds. This same principle can guide us to some extent in our Galilean case. In our case, let's consider Galilean boost by parameter $\beta^i$ and consider 
\be{}
\delta_{\beta} \Lambda_{1,2} = \beta^i[B_i, \Lambda_{1,2}] =t\, \beta^i \partial_i \Lambda_{1,2}. 
\ee
The caveat here is that we could have added a $SO(D)$ vector to the transformation as this is allowed by the representation. One can further go on restricting this by other consistency conditions like invariance of equation of motion. But again, that's a choice, like the one we had made by keeping only scalars and vectors in our $SO(D)$ multiplet of field content of GED. Therefore we have purposefully did not add a vector $\Lambda_{(1,2),i}$ in the `$\Lambda_{1,2}$ multiplet'. Otherwise we should have written \eqref{gt} in a way, such that
$$A_i \rightarrow A_i+ \partial_i \Lambda_2 + \partial_t \Lambda_{2,i}.$$
Now one can implement the boost-gauge commutation. A short analysis starting from this enforces that in the Electric limit $\partial_t \Lambda_1=0=\partial_i \Lambda_1$. Hence, $A_t$ does not gauge transform, while $A_i$ does. More tests, ie gauge invariance of the equations of motion then shows that no further constraints are put on $\Lambda_2$. Similar analysis in magnetic limit shows $\partial_i\Lambda_1=\partial_i\Lambda_2$, hence $A_t$ and $A_i$ transform in same way. No further restriction is put by equations of motion.

These results derived completely from a set of arguments intrinsic to the Galilean theory are consistent with the ones found by the ones found by scaling appropriately the relativistic rules \refb{elga}, \refb{maga}.

\subsection{Relativistic conformal invariance of Yang-Mills theory}
We wish to remind the reader of the classical conformal invariance of Yang-Mills theory. 
In order to set some notation, here are some details of the conformal algebra in $D=4$. 
\begin{flalign}
\mbox{Poincare generators:} \quad \tilde P_i= \p_i, \tilde H=-\p_t, \tilde J_{i j} = x_{[i} \p_{j]}, \tilde B_i = x_i \p_t + t \p_i \\
\mbox{Conformal generators:} \quad \tilde D = -x \cdot \p, \ \tilde K_\mu =  - (2 x_\mu (x \cdot \p)  - (x \cdot x) \p_\mu)
\end{flalign}
The conformal algebra in $D$ dimensions is isomorphic to $\mathfrak{so}(D,2)$. Let us indicate a few important commutation relations below, so that the differences with the GCA \refb{GCA} is apparent:
\be{rel-al}
[\tilde P_i, \tilde B_j]= - \delta_{ij} \tilde H, \, \, [\tilde B_i, \tilde B_j] =  \tilde J_{ij}, \, \, [\tilde K_i, \tilde B_j] = \delta_{ij} \tilde K, \, \, [\tilde K_i, \tilde P_j] = 2 \tilde J_{ij} + 2 \delta_{ij} \tilde D. 
\ee
The right hand side of all these commutators are zero in the f-GCA, while all other commutators stay the same. 
Now, following \cite{Jackiw:2011vz}, we describe the conformal transformations of fields. Poincare transformations of a multi-component field $\Phi$:
\be{}
{\delta}_\mu^{\mbox{\tiny{P}}} \, \Phi (x) = \p_\mu \Phi (x), \quad {\delta}_{\mu \nu}^{\mbox{\tiny{L}}} \, \Phi (x) = (x_\mu \p_\nu - x_\nu \p_\mu + \Sigma_{\mu\nu}) \Phi(x)
\ee
Transformations under scaling takes the following form:
\be{}
{\delta}^{\mbox{\tiny{D}}}  \, \Phi (x) = (x \cdot \p + {\tilde{\Delta}}) \Phi
\ee
where ${\tilde{\Delta}}$ is the scaling dimension of the field $\Phi$ and to make the kinetic term of the corresponding action scale invariant one chooses
\be{De}
{\tilde{\Delta}} = \frac{D-2}{2}.
\ee
The transformation under special conformal transformation specialised to the case of primary fields:
\be{}
{\delta}^{\mbox{\tiny{K}}}_\mu  \, \Phi (x) = \{ 2 x_\mu (x \cdot \p) - x^2 \p_\mu + 2 {\delta} x_\mu - 2 x^\nu \Sigma_{\nu \mu} \}  \, \Phi (x).
\ee

We now consider Yang-Mills theory in D-dimensional spacetime. The theory is best expressed in terms of a field strength 
$$F^a_{\mu\nu} = \p_\mu A^a_\nu - \p_\nu A^a_\mu + g f_{abc} A^b_\mu A^c_\nu$$ 
where $A^a_\mu$ is the fundamental dynamic variable, the gauge field. The label $a$ is the colour index and $f_{abc}$ are the structure constants of the underlying gauge group with generators $T^a$ following the algebra: 
$$[T^a, T^b] = f_{abc} T^c.$$
The equations of motion
\be{}
 \p^\mu F^a_{\mu \nu} + g  f_{abc}  A^{b \mu} F^c_{\mu\nu} =0
\ee
can be derived from the well-known Lagrangian
\be{L}
\mathcal{L} = - \frac{1}{4} \text{Tr} \ F_{\mu \nu} F^{\mu \nu}
\ee
While the above Lagrangian is manifestly invariant under Poincare transformations:
\bes \label{poinc}
\bea{}
&&{\delta}_\mu^{\mbox{\tiny{P}}} \, A^a_{\nu} (x) = \p_\mu A^a_{\nu} (x) \\ &&{\delta}_{\mu \nu}^{\mbox{\tiny{L}}} \, A^a_{\rho} (x) = (x_\mu \p_\nu - x_\nu \p_\mu)A^a_{\rho}(x) + \eta_{\rho \mu} A^a_{\nu}(x)-\eta_{\rho \nu} A^a_{\mu}(x),
\eea
\ees
scale and special conformal transformations act non-trivially on it. In terms of the field variable $A_\mu$, which we treat as a vector primary field, the transformations are the following:
\bea{}
&& {\delta}^{\mbox{\tiny{D}}}A^{a}_{\mu}(x)=(x^{\nu}\partial_{\nu}+\Delta)A^{a}_{\mu}\\
&& {\delta}^{\mbox{\tiny{K}}}_{\sigma} A_{\nu}^{a}(x)=(2x_{\sigma}x_{\mu}-\eta_{\sigma\mu} x^{2})\partial^{\mu}A_{\nu}^{a}+(D-2) x_{\sigma}A_{\nu}^{a}-2x_{\nu}A_{\sigma} ^{a}+2\eta_{\sigma\nu}x^{\mu}A_{\mu}^{a} 
\eea 
where as before in \refb{De}, we have $\Delta=\frac{D-2}{2} $. To examine the symmetry of the equations of motion, we need the transformation of the field strength. Under dilatations, we have:
\be{} 
{\delta}^{\mbox{\tiny{D}}} F^{a}_{\mu\nu}(x)=(x^{\rho}\partial_{\rho}+\Delta+1)F^{a}_{\mu\nu}+gf^{abc}(\Delta-1)A^{b}_{\mu}A_{\nu}^{c},
\ee
while under special conformal transformations, the field strength transforms as:
\bea{} 
{\delta}^{\mbox{\tiny{K}}}_{\sigma}F_{\mu\nu}^{a}(x)=(2x_{\sigma}x_{\mu}-\eta_{\sigma\mu} x^{2})\partial^{\mu}F_{\mu\nu}^{a}+Dx_{\sigma}F_{\mu\nu}^{a}+2\eta_{\sigma\mu}
x^{\tau}F_{\tau\nu}^{a}
+2\eta_{\sigma\nu}x^{\rho}F_{\mu\rho}^{a}\non\\-2x_{\mu}F_{\sigma\nu} ^{a}
-2x_{\nu}F_{\mu\sigma}^{a}+(D-4)\left[(\eta_{\sigma\mu}A_{\nu}^{a}-\eta_{\sigma\nu}A_{\mu}^{a})
+gf^{abc}x_{\sigma}A^{b}_{\mu}A^{c}_{\nu}\right].
\eea

We now wish to examine the action of the various transformations on the equations of motion of the YM theory. The Poincare transformations obviously leave the EOM invariant. We first check for invariance under dilatations. 
\bea{}
\delta^{\mbox{\tiny{D}}} \left[ \p^\mu F^a_{\mu \nu} + g  f_{abc}  A^{b \mu} F^c_{\mu\nu} \right] && = \partial^{\mu}\delta^{\mbox{\tiny{D}}} F^{a}_{\mu\nu}+gf^{abc}\delta^{\mbox{\tiny{D}}}(A^{\mu b}F^{c}_{\mu\nu})\non \\ 
= gf^{abc} && (\Delta-1)\left[\partial^{\mu}(A^{b}_{\mu}A^{c}_{\nu})+A^{\mu b}(F^{c}_{\mu\nu}+gf^{cde}A^{d}_{\mu}A^{e}_{\nu})\right]. 
\eea
This is zero only for $\Delta =1$ which indicates that Yang-Mills theory is scale invariant only in $D=4$. This is a departure from Maxwellian electrodynamics, which is scale invariant in all dimensions. Checking for the transformations of equations of motion under special conformal transformations, we find
\bea{eom8}
&&\partial^{\alpha}\delta^{K}_{\sigma}F^{a}_{\alpha\beta}+gf^{abc}\delta^{K}_{\sigma}(A^{\alpha b}F^{c}_{\alpha\beta}) = (D-4) \bigg[ F^{a}_{\sigma \beta}+(\partial_{\sigma}A^{a}_{\beta}-\eta_{\sigma \beta}\partial^{\alpha}A^{a}_{\alpha})+ \non\\
g f^{abc} && \bigg\{2A^b_\sigma A^{c}_{\beta}-\eta_{\sigma \beta}A^{\alpha b}A^{c}_{\alpha} +\eta_{\sigma \rho} x^{\rho}\left(\partial^{\alpha}(A^{b}_{\alpha}A^{c}_{\beta})+A^{\alpha b}F^{c}_{\alpha\beta}+gf^{cde}A^{\alpha b}A^{d}_{\alpha}A^{e}_{\beta}\right)\bigg\} \bigg] \non
\eea 
which implies that the EOM are also invariant in $D=4$ under special conformal transformations. Thus we see that classical Yang-Mills theories are invariant under the full conformal group in $D=4$. We have checked the invariance of the EOM, but as is well known, this can be checked also at the level of the action of the theory. The process we elucidated above is useful for non-relativistic theories as we have said before, since we don't (yet) have an action formulation for the theories we consider later in this paper.  

\newpage

\section{Galilean Yang Mills: the $SU(2)$ story} \label{su2section}
In this section, we will work out the details of the non-relativistic limit of the simplest non-Abelian Yang-Mills theory, {\it{viz.}} one with $SU(2)$ gauge symmetry. This would be the first example of an {\em{interacting}} GCFT in $D>2$. 

The first non-trivial aspect of the generalisation of the gauge group from $U(1)$ to $SU(2)$ is the existence of skewed limits, over and above the Electric and Magnetic limits in the Galilean Electrodynamics. This is because we now have three different gauge fields in the game instead of just one and each pair $(A_0, A_i)$ can have electric and magnetic limits. This leads to four distinct limits instead of two in the case of the $U(1)$ theory. We will consider these one by one. For each sector, we would state the scaling, construct the equations of motion, look at gauge invariance and then check the symmetries of the equations of motion. In the previous section, we addressed gauge invariance of the Galilean Electrodynamics in two separate ways, one as a limit and the other an intrinsic analysis. In this section, we will only look at the limiting construction for gauge invariance. We come back to the intrinsic analysis for the general $SU(N)$ analysis that we present in the next section. 

\subsection{EEE: Electric sector} \label{eee}
We begin by looking at the ``vanilla" electric limit, where all the gauge fields transform in the same way. 

\paragraph{{Scaling:}} As stated above all the gauge fields transform in the same way. 
\be{}
A^a_t \to A^a_t, \quad A^a_i \to \e A^a_i.
\ee

\smallskip

\paragraph{{Equations of motion:}} We apply the above scalings on the equations of motion of Yang Mills theory to obtain the EOM for the electric limit. 
\be{eomEEE}
(\partial.\partial) A^{a}_{i}-\partial^{j}\partial_{i}A^{a}_{j}+\partial_{t}
\partial_{i}A_{t}^{a}+g\varepsilon^{abc}A^{b}_{t}\partial_{i}A^{c}_{t}=0, \quad \partial^{i}\partial_{i} A^{a}_{t}=0
\ee

\smallskip

\paragraph{{Gauge invariance:}} We now look at the remnants of gauge invariance in this limit. For relativistic Yang Mills theory with an arbitrary gauge group, the theory is invariant under gauge transformations of the form
\be{ymga}
A_{\mu}^{a} \rightarrow A_{\mu}^{a}+\frac{1}{g}\partial_{\mu}\alpha^{a}+f^{abc}A^{b}_{\mu}\alpha^{c} 
\ee
where $f^{abc}$ are the structure constants of the underlying gauge algebra and $\alpha^{a}$ are arbitrary functions of spacetime. Generalising our strategy for the Galilean Electrodynamics which was detailed in Sec \refb{gauge}, we work out the gauge invariance in this limit of the YM theory. We apply the scaling \refb{eomEEE} on the equation \refb{ymga} and check what scaling of $\alpha^a$ keeps the equation finite. We find that the scaling needs to be 
\be{}
\alpha^a \to \e^2 \alpha^a. 
\ee
The non-relativistic version of gauge invariance in this limit reads:
\be{ggel} 
A_t^{a} \rightarrow A_t^{a}, \quad A_i^{a}  \rightarrow A_i^{a}+ \frac{1}{g}\partial_i \alpha^{a}. 
\ee
It can be checked that the EOM \refb{eomEEE} are invariant under the above transformations. Note that the vanilla electric limit leads to gauge invariance which does not retain its non-Abelian nature. The reason behind this would become clear when we are looking at the general structure of gauge invariance for a $SU(N)$ theory. 

\bigskip

\paragraph{{Finite Galilean conformal symmetry of EOM:}} We saw that the relativistic Yang-Mills equations of motion were invariant under the full relativistic conformal group in $D=4$. A scaling limit of these equations lead to \refb{eomEEE} and the same limit on the conformal group lead to the GCA. It is thus expected that the electric EOM would display invariance under the GCA. We now explicitly verify this expectation following the procedure reviewed earlier in Sec.\refb{inveom}. 

The scale-spin representations of GCA as stated before is determined by the set $\{ \Delta, \Sigma, r, s\}$. We are dealing with a set of scalar and vector primaries $A_t^a, A_i^a$.  For each gauge copy $a$, we would have a specific $(r, s)$ and hence these are vector valued and will be called $(r^a, s^a)$. For the present (EEE) case, we have 
\be{rsEEE}
\{ (r_1, s_1), (r_2, s_2), (r_3, s_3) \} = \{ (-1, 0), (-1, 0), (-1, 0) \}.
\ee
We now have the ingredients of the representation theory to address the main question at hand: the symmetries of the equations of motion. We would first consider the transformation of the EOM \refb{eomEEE} under dilatations. 
\bea{} 
(\partial.\partial) \lb D,A_{i}^{a} \rb - \partial_i  \partial^j \lb D,A_{j}^{a} \rb + \partial_t \partial_ i \lb D, A_{t}^{a} \rb +g\varepsilon^{abc}[D,A^{b}_{t} \partial_{i}A_{t}^{c}] &=& \frac{1}{2} (D-4)g \varepsilon^{abc}A^{b}_{t}\p_{i}A^{c}_{t}. \non\\
\partial^{j}\partial_{j} [ D,A_t^{a} (t,x) ] &=& 0.
\eea
We find that the equations are invariant under the dilatation operator in $D=4$. Now the more non-trivial check is for invariance under the Galilean special conformal transformations, $K$ and $K_i$.  Invoking \refb{rsEEE}, the invariance of the second equation of \refb{eomEEE} is immediate: 
\be{}
\partial^{j}\partial_{j} \lb K, A_t^{a} (t,x) \rb = 0, \quad \partial^{j}\partial_{j} \lb K_i ,A_t^{a} (t,x) \rb = 0 
\ee
Below we check the transformation of the first equation of \refb{eomEEE}:  
\bea{}  
\partial . \partial \lb K,A_{i}^{a} \rb - \partial_i  \partial^j \lb K,A_{j}^{a} \rb + \partial_t \partial_ i \lb K, A_{t}^{a} \rb +g\varepsilon^{abc}[K,A^{b}_{t}(\partial_{i}A_{t}^{c})] \non\\ 
= (D-4)[-\p_{i}A^{a}_{t}+ g\varepsilon^{abc}tA^{b}_{t}\p_{i}A^{c}_{t}];  \\
\non\\
\partial . \partial \lb K_{l},A_{i}^{a} \rb - \partial_i  \partial^j \lb K_{l},A_{j}^{a} \rb + \partial_t \partial_ i \lb K_{l}, A_{t}^{a} \rb +g\varepsilon^{abc}[K_{l},A^{b}_{t}(\partial_{i}A_{t}^{c})]=0.
\eea
We see that the EOM are invariant under $K_i$ in all dimensions, but only invariant under $K$ in $D=4$. Hence we have proved what was expected, viz. the equations of motion of Galilean Yang Mills theory in the Electric limit are invariant under the finite GCA in $D=4$. This was to be expected given that the relativistic theory was invariant under the conformal group in $D=4$. 

\bigskip

\paragraph{{Infinite Galilean conformal symmetry of EOM:}} The very non-trivial part of our analysis in this particular limit is the proof that the set of equations \refb{eomEEE}, unlike their relativistic counterparts, actually exhibit an infinite dimensional symmetry. This is the symmetry of the extended GCA \refb{GCA}. We will use the knowledge of the representation theory discussed earlier, specifically \refb{inf} to check for the transformation of \refb{eomEEE} under the infinite algebra. Implicit in this analysis would be the knowledge of the set $(r^a, s^a)$ \refb{rsEEE}. We first check the transformations under $M^{(n)}_{i}$:
\bea{}
&&(\partial \cdot \partial) [M_{l}^{(n)},A_{i}^{a}] - \partial_i  \partial^j [M_{l}^{(n)},A_{j}^{a}] + \partial_t \partial_ i [M_{l}^{(n)}, A_{t}^{a} ]  +g\varepsilon^{abc}[M_{l}^{(n)},A^{b}_{t} \partial_{i}A^{c}_{t}] = 0 \non\\
&&(\partial \cdot \partial) \lb M_{i}^{(n)} ,A_{t}^{a}  \rb = 0
\eea
So we see that the Electric EOM of Galilean Yang-Mills theory has an infinite dimensional symmetry under all the $M_i^{(n)}$ and this is true in all dimensions. We now move on to transformations of the equations under  $L^{(n)}$'s. 
\bea{}
(\partial \cdot \partial) \lb L^{(n)} ,A_{i}^{a} \rb &-& \partial_i  \partial^j \lb L^{(n)},A_{j}^{a} \rb + \partial_t \partial_ i \lb L^{(n)}, A_{t}^{a} \rb  +g\varepsilon^{abc}[L^{(n)},A^{b}_{t}\partial_{i}A^{c}_{t}] \non\\ &=& \frac{1}{2}(D-4)(n+1) \left(-nt^{n-1}\partial_{i}A^{a}_{t}+g\varepsilon^{abc}t^{n}A_{t}^{b}
\partial_{i}A^{c}_{t} \right) \\
(\partial \cdot \partial) \lb L^{(n)},A_{t}^{a}  \rb &=& 0
\eea
Hence we have shown that the EOM are also invariant under all $L^{(n)}$ in $D=4$. Hence we have invariance of the equations of motion of the Electric limit of Galilean Yang Mills theory under the full infinite dimensional GCA in $D=4$. This is a limit which contains interactions, as is evident from \refb{eomEEE} and hence constitutes the first example of an interacting GCFT in $D>2$. The following subsections will reveal similar results. 
The careful reader may notice that we have not explicitly shown the invariance of the EOM under negative modes. The invariance under the negative modes indeed does hold here and in all subsequent sub-cases to be discussed below. We refer the reader to Appendix A for a treatment of these modes.

\subsection{EEM: Skewed sector 1} \label{eem}
We now turn our attention to the first skewed limit, where two of the pairs $(A_t^a, A_i^a)$ scale electrically and the remaining pair scales magnetically. 

\paragraph{Scaling:} The gauge fields transform according to: 
\be{sceem}
A^{1,2}_t \to A^{1,2}_t, \quad A^{1,2}_i \to \e A^{1,2}_i; \qquad A^3_t \to \e A^3_t, \quad A^3_i \to A^3_i.
\ee
There is obviously no difference in which pair scales magnetically and which two scale electrically. Scaling $A^1$ magnetically and $A^{2,3}$ electrically would lead to the same results. 
Here we will stick to the above scaling \refb{sceem}. 
\smallskip

\paragraph{Equations of Motion:}
The equations of motion in this limit are strangely devoid of interaction terms. They are given by: 
\bes \label{EEMeom}
\bea{}
&& \partial^{i}\partial_{i} A_{t}^{1,2}=0,\quad \quad
 \partial^{i}\partial_{i}A_{j}^{1,2}
 -\partial^{i}\partial_{j}A_{i}^{1,2}
 +\partial_{t}\partial_{j}A_{t}^{1,2}=0\\
&& \partial^{i}\partial_{i} A_{t}^{3}-\partial^{i}\partial_{t}A_{i}^{3}=0,\quad
\partial^{i}(\partial_{i}A_{j}^{3}-\partial_{j}A_{i}^{3})=0
\eea
\ees
The absence of interactions in this limit would be more clear when we generalise our analysis to the $SU(N)$ theory in the next section and look at the structure of the equations of motion in a general skewed limit consisting of an arbitrary number of electric and magnetic legs.  
\smallskip

\paragraph{Gauge Invariance:}
To get the gauge transformations of the fields for this limit, we shall consider the scaling \refb{sceem} and in addition, the $\alpha^{a}$'s should transform as  
\be{IIIgi}
\alpha^{1,2}\rightarrow \epsilon^{2} \alpha^{1,2}, \quad \alpha^{3}\rightarrow \epsilon \alpha^{3}.   
\ee
Gauge invariance in this limit reads
\bes
\bea{} 
A^{1,2}_{t}\rightarrow A^{1,2}_{t},\quad
 A^{1,2}_{i}\rightarrow A^{1,2}_{i}+ \frac{1}{g}\partial_{i}\alpha^{1,2} \\
A^{3}_{t}\rightarrow A^{3}_{t}+\frac{1}{g}\partial_{t}\alpha^{3},\quad
 A^{3}_{i}\rightarrow A^{3}_{i}+ \frac{1}{g}\partial_{i}\alpha^{1}
 \eea
 \ees
It is easy to check that the equations of motion (\ref{EEMeom}) remain invariant under the above transformations. Given that there are no interaction terms in the equations of motion, it is not a surprise that the gauge invariance in this limit does not contain any hint of the non-Abelian nature of the parent relativistic theory.  

\smallskip

\paragraph{Finite Galilean Conformal symmetry of EOM:}

We now wish to check the symmetries of the equations of motion \refb{EEMeom}. For this, the first information we need is the set of vectors $\{ \vec{r}, \vec{s} \}$ which fix the details of the representation theory. This is given by
\be{rsEEM}
\{ (r_1, s_1), (r_2, s_2), (r_3, s_3) \} = \{ (-1, 0), (-1, 0), (0, -1) \}.
\ee
We will use this directly in the calculations. The fact that there are no interaction terms makes most of the calculations in this limit immediate. Checking for scale invariance is straight-forward as is checking for the invariance under $K_i$. We shall not bother the reader with details of these. We only present below the invariance under $K$ of the second and third equations of \refb{EEMeom}. 
\bes
\bea{}
&& \partial^{i}\partial_{i}[K,A_{j}^{1,2}]-\partial^{i}\partial_{j}[K,A_{i}^{1,2}]+\partial_{t}\partial_{j}[K,A_{t}^{1,2}]=-(D-4)\partial_{j}A_{t}^{1,2} \\
&& \partial^{i}\partial_{i} [K,A_{t}^{3}]-\partial^{i}\partial_{t}[K,A_{i}^{3}]=-(D-4)\partial_{i}A_{i}^{3}
\eea \ees
So we see that these equation are invariant only in $D=4$, as expected. The other two equations of \refb{EEMeom} are invariant in all dimensions. 

\smallskip

\paragraph{Infinite Galilean Conformal symmetry of EOM:}
Checking for the infinite dimensional invariance is also straight-forward using \refb{rsEEM} and \refb{inf}.  Invariance under $M^{(n)}_i$ is immediate and again we display the check for the second and third equations of \refb{EEMeom} under a general $L^{(n)}$: 
\bea{} \partial.\partial [L^{(n)},A_{t}^{3}]-\partial^{i}\partial_{t}[L^{(n)},A_{i}^{3}]=-\frac{1}{2}(D-4)n(n+1)t^{n-1}\partial_{i}A_{i}^{3}\eea
\bea{}
\partial^{i}\partial_{i}[L^{(n)},A_{j}^{1,2}]-\partial^{i}\partial_{j}[L^{(n)},A_{i}^{1,2}]+\partial_{t}\partial_{j}[L^{(n)},A_{t}^{1,2}]\non\\=-\frac{1}{2}(D-4)n(n+1)t^{n-1}\partial_{j}A_{t}^{1,2}
\eea
So we find that the equations of motion in this limit are also invariant under all the generators of the infinite dimensional GCA. 

\bigskip

\subsection{EMM: Skewed sector 2} \label{emm}
The second skewed limit, where one pair scales electrically and the remaining two pairs scales magnetically, turns out to be the most interesting of all the four limits that exist in the $SU(2)$ Galilean YM theory. Again the reason behind why this is the case will be better understood when we address the general $SU(N)$ construction in the next section. For the moment, let us present the details of the case at hand.  

\paragraph{Scaling:} We choose $A^1$ to scale electrically and the others to scale magnetically: 
\be{II}
A_{t}^{1}\to A_{t}^{1}, \quad A_{i}^{1}\to \epsilon A_{i}^{1}; \quad A_{t}^{2,3}\to \epsilon A_{t}^{2,3}, \quad A_{i}^{2,3}\to A_{i}^{2,3}.
\ee

\smallskip

\paragraph{Equations of Motion:}
In this scaling limit, we have more involved equations of motion than any of the other cases discussed. The reason for this would become apparent when we look at the general analysis in the next section. They are given below: 
\bes\label{EMMeom}
\bea{}
\partial^{i}(\partial_{i}A_{j}^{1}-\partial_{j}A_{i}^{1})
+\partial_{t}\partial_{j}A_{t}^{1} +g \partial^{i}(A^{2}_{i}A^{3}_{j}-A^{3}_{i}A^{2}_{j}) &&  \non\\
+ g A^{2}_{i}(\partial_{i}A^{3}_{j}-\partial_{j}A^{3}_{i})+ g A^{3}_{i}(\partial_{j}A^{2}_{i}-\partial_{i}A^{2}_{j}) &=& 0  \label{eom1}
\\
\non\\
 \partial^{i}\partial_{i} A^{1}_{t}=0, \quad
\partial^{i}(\partial_{i}A_{j}^{2,3}-\partial_{j}A_{i}^{2,3})&=&0 \label{eom2}\\
\non\\
\partial^{i}\partial_{i} A_{t}^{3}-\partial^{i}\partial_{t}A_{i}^{3} -2g A_{i}^{2}\partial^{i}A_{t}^{1}
-gA_{t}^{1}\partial^{i}A_{i}^{2}&=&0 \label{eom3}\\ 
\partial^{i}\partial_{i} A_{t}^{2}-\partial^{i}\partial_{t}A_{i}^{2} +2g A_{i}^{3}\partial^{i}A_{t}^{1}+gA_{t}^{1}
 \partial^{i}A_{i}^{3}&=&0 \label{eom4}
\eea
\ees
Note that here we have used the structure constant of $SU(2)$, i.e. $f_{abc}= \varepsilon_{abc}$ and put in the values of $\varepsilon_{abc}$ (which are $\pm 1$) for different permutations directly into the equations.  

\smallskip

\paragraph{Gauge Invariance:}

For the gauge transformation of the potentials we shall again consider the $A^{1}$ in the electric limit and $A^{2,3}$ in the magnetic limit. Here, we have $\alpha^{a}(a=1,2,3)$ which will also get scaled in accordance with the $A$'s being in one of the two limits. 
\be{IIgi}  \alpha^{1}\rightarrow \epsilon^{2} \alpha^{1}, \hspace{.2cm}
\alpha^{2,3}\rightarrow \epsilon \alpha^{2,3}
\ee
To obtain the gauge transformations associated to this limit, we shall use (\ref{IIgi}) in order to get the desired transformations of the fields as below. Here we have again used the explicit values of the structure constants of $SU(2)$.
\bes
\bea{} 
A^{1}_{t}\rightarrow A^{1}_{t},\quad A^{1}_{i}\rightarrow A^{1}_{i}+ \frac{1}{g}\partial_{i}\alpha^{1}+A^{2}_{i}\alpha^{3}-A^{3}_{i}\alpha^{2}; \\
A^{3}_{t}\rightarrow A^{3}_{t}+\frac{1}{g}\partial_{t}\alpha^{3}+A^{1}_{t}\alpha^{2},\quad A^{3}_{i}\rightarrow A^{3}_{i}+ \frac{1}{g}\partial_{i}\alpha^{3};\\
 A^{2}_{t}\rightarrow A^{2}_{t}+\frac{1}{g}\partial_{t}\alpha^{2}-A^{1}_{t}\alpha^{3},\quad A^{2}_{i}\rightarrow A^{2}_{i}+ \frac{1}{g}\partial_{i}\alpha^{2}.
\eea
\ees
It can be checked that the above leave the equations of motion (\ref{EMMeom}) invariant. It is important to note that in this sector of Galilean YM, we have non-Abelian structure in our gauge invariance as well as in the equations of motion, making this the most interesting of the limits considered in this explicit example.

\smallskip

\paragraph{Finite Galilean Conformal symmetry of EOM}
Finding the symmetries of the equations of motion \refb{EMMeom} is our present goal. To this end, just like in the previous case, we will need the set of vectors $\{ \vec{r}, \vec{s} \}$ that fix the details of the representation theory in this particular sector. This is given by
\be{rsEMM}
\{ (r_1, s_1), (r_2, s_2), (r_3, s_3) \} = \{ (-1, 0), (0, -1), (0, -1) \}. 
\ee
We will use this information in the calculations directly. The calculations are very similar to the ones carried out earlier. So for this subsection, we shall only display the invariance of the most interesting of the equations of motion, viz. \refb{eom1}. Transformation under $D$:
\bea{}
&&\p ^i[D, (\p_i A_j^1 - \p_j A_i^1)]+\partial_{t}\partial_{j}[D,A_{t}^{1}] +g \partial^{i}[D, (A^{2}_{i}A^{3}_{j}-A^{3}_{i}A^{2}_{j})]\non \\
&&+ g [D, A^{2}_{i}(\partial_{i}A^{3}_{j}-\partial_{j}A^{3}_{i})]+ g [D, A^{3}_{i}(\partial_{j}A^{2}_{i}-\partial_{i}A^{2}_{j})]  \\
&=&\frac{1}{2}(D-4)[2g(A_{i}^{2}\partial^{i}A_{j}^{3}-A_{i}^{3}\partial^{i}A_{j}^{2}) +g(A_{j}^{3} \partial^{i}A_{i}^{2}-A_{j}^{2}\partial_{i}A_{i}^{3}-A^{i2}\partial_{j}A_{i}^{3}
+A^{i3}\partial_{j}A_{i}^{2})] \non
\eea
It is evident that the equation is invariant only in $D=4$ under dilatations. Invariance under $K_i$ is immediate. Under $K$, we have
\bea{}
\delta^K \refb{eom1} =(D-4)[-\partial_{j}A_{t}^{1}+2gt(A_{i}^{2}\partial^{i}A_{j}^{3}-A_{i}^{3}\partial^{i}A_{j}^{2})+gt(A_{j}^{3}
\partial^{i}A_{i}^{2}\non\\-A_{j}^{2}\partial_{i}A_{i}^{3}-A^{i2}\partial_{j}A_{i}^{3}+A^{i3}\partial_{j}A_{i}^{2})]
\eea
Again, it is clear that the equation exhibits invariance under the special conformal transformations only for $D=4$. So we have shown the invariance of \refb{eom1} under the finite GCA. The invariances of the other equations are straight forward and one can look at the transformations of the equations under the infinite GCA that we discuss next, plug in the appropriate values of $n$ in $L^{(n)}$, $M^{(n)}_i$ and obtain the relevant formulae for the other equations. 
\smallskip

\paragraph{Infinite Galilean Conformal symmetry of EOM:}
We want to extend our analysis for the infinite number of modes of galilean conformal algebra. we would be following the same procedure as above and will see the invariance (under I-GCA) of the equations of motion.
Checking for Eq \refb{eom2}: 
\bes\bea{} 
\partial^{i}\partial_{i} [L^{(n)},A^{1}_{t}]=0, \hspace{.4 cm}
\partial^{i}\partial_{i}[L^{(n)},A_{j}^{2,3}]-\partial^{i}\partial_{j}[L^{(n)},A_{i}^{2,3}]=0\\
\partial^{i}\partial_{i} [M_{l}^{(n)},A^{1}_{t}]=0, \hspace{.4 cm}
\partial^{i}\partial_{i}[M_{l}^{(n)},A_{j}^{2,3}]-\partial^{i}\partial_{j}[M_{l}^{(n)},A_{i}^{2,3}]=0\eea\ees
Checking for Eq \refb{eom3}:
\bea{}
\partial^{i}\partial_{i} [L^{(n)},A_{t}^{3}]-\partial^{i}\partial_{t}[L^{(n)},A_{i}^{3}] -g[L^{(n)},A_{t}^{1}\partial^{i}A_{i}^{2}+2 A^{i2}\partial_{i}A_{t}^{1}]=\non
\\ -\frac{1}{2}(D-4)(n+1)[nt^{n-1}\partial_{i}A_{i}^{3}+gt^{n}(A_{t}^{1}\partial^{i}A_{i}^{2}
+2A_{i}^{2}\partial^{i}A_{t}^{1})]\\
\partial^{i}\partial_{i} [M_{l}^{(n)},A_{t}^{3}]-\partial^{i}\partial_{t}[M_{l}^{(n)},A_{i}^{3}] -g[M_{l}^{(n)},A_{t}^{1}\partial^{i}A_{i}^{2}+2 A^{i2}\partial_{i}A_{t}^{1}]=0
\eea
Checking for Eq \refb{eom4}: 
\bea{}
\partial^{i}\partial_{i}[L^{(n)},A_{t}^{2}]-\partial^{i}\partial_{t}[L^{(n)},A_{i}^{2}]+
g[L^{(n)},A_{t}^{1}\partial^{i}A_{i}^{3}+2 A^{i3}\partial_{i}A_{t}^{1}]=\non\\
\frac{1}{2}(D-4)(n+1)[-nt^{n-1}\partial_{i}A_{i}^{2}+gt^{n}(A_{t}^{1}\partial^{i}A_{i}^{3}+2 A_{i}^{3}\partial^{i}A_{t}^{1})]\\
\partial^{i}\partial_{i}[M_{l}^{(n)},A_{t}^{2}]-\partial^{i}\partial_{t}[M_{l}^{(n)},A_{i}^{2}]+
g[M_{l}^{(n)},A_{t}^{1}\partial^{i}A_{i}^{3}+2 A^{i3}\partial_{i}A_{t}^{1}]=0
\eea
Following on the same steps for \refb{eom1} gives (here we don't write the left hand side of the equation explicitly): 
\bea{}
 \delta^{M^{(n)}_i} \refb{eom1} &=& 0 \\
\delta^{L^{(n)}} \refb{eom1} &=& \frac{1}{2}(D-4)(n+1)[-nt^{n-1}\partial_{j}A_{t}^{1}+2gt^{n}(A_{i}^{2}\partial^{i}A_{j}^{3}-A_{i}^{3}\partial^{i}A_{j}^{2})\non\\&&+gt^{n}(A_{j}^{3}\partial^{i}A_{i}^{2}-A_{j}^{2}\partial_{i}A_{i}^{3}-A^{i2}\partial_{j}A_{i}^{3}
 +A^{i3}\partial_{j}A_{i}^{2})]
\eea
We have thus shown the invariance of the equations of motion of this skewed sector under the infinite dimensional GCA in $D=4$.

\smallskip

\subsection{MMM: Magnetic sector} \label{mmm}
Perhaps the most uninteresting sector of the $SU(2)$ theory is the ``vanilla" magnetic sector. This is a sector that does not exhibit any interactions in the equations of motion as well as in gauge invariance and it just is three copies of the $U(1)$ magnetic sector. We shall just illustrate the equations of motion and the gauge invariance for completeness and refer the reader to \cite{Bagchi:2014ysa} for the checking of the symmetries of the equations. 

\paragraph{Scaling:} All fields scale in the same way.
\be{Ml}
A_{t}^{a}\rightarrow \epsilon A_{t}^{a}, \quad A_{i}^{a}\rightarrow  A_{i}^{a}
\ee

\paragraph{Equations of Motion:} As stated before, there are no interaction terms in the equations of motion which reduce to copies of the ones of the magnetic sector of Galilean electrodynamics:
\be{MMMeom}
 \partial^{i}\partial_{i}A^{a}_{t}-\partial^{i}
\partial_{t}A^{a}_{i}=0, \quad
\partial^{j}\partial_{j} A^{a}_{i}-\partial^{j}\partial_{i}A^{a}_{j}=0.
\ee

\smallskip

\paragraph{Gauge Invariance:} The gauge transformation for magnetic limit can be found by taking the scaling of potentials and $\alpha^{a}$ as 
\bea{} \alpha^{a}\rightarrow \epsilon \alpha^{a}\eea
Even gauge invariance does not exhibit any non-Abelian structure. Using the limit on (\ref{ymga}), the equation becomes
\bea{} A_i^{a}  \rightarrow A_i^{a}+ \frac{1}{g}\partial_i \alpha^{a},\hspace{.4 cm}
A_t^{a}  \rightarrow A_t^{a}+ \frac{1}{g}\partial_t \alpha^{a}\eea
Invariance of equations of motion can be easily checked by plugging the above back into the equations \refb{MMMeom}.

\smallskip

\paragraph{Galilean Conformal symmetry of EOM:} We shall not be explicitly writing these down here as the checks are the same as that for the case of the magnetic sector of Galilean Electrodynamics. The only required input is the set of vectors $\{ \vec{r}, \vec{s} \}$ that fix the details of the representation theory in this particular sector. This is given by
\be{rsMMM}
\{ (r_1, s_1), (r_2, s_2), (r_3, s_3) \} = \{ (0, -1), (0, -1), (0, -1) \}. 
\ee
The rest of the analysis is straight-forward and identical to the electrodynamics case. The interested reader is referred to \cite{Bagchi:2014ysa} (section 5.2, pages 18- 19). The upshot is that the equations of motion in this sector, like the other sectors, is invariant under all the modes of the infinite dimensional GCA.

\bigskip

We have thus looked at Galilean $SU(2)$ Yang-Mills theory, discovered that there are four distinct sectors within the Galilean theory, all of which exhibit classical Galilean conformal symmetry in $D=4$. This is the first example of an {\em{interacting}} GCFT in $D>2$. 

\newpage
  
\section{Galilean Yang Mills: $SU(N)$ analysis}
Motivated by the success of the first non-trivial interacting (non-Abelian) gauge theory, {\it{viz.}} {the Galilean $SU(2)$ YM discussed in the the last section}, it is natural to probe into the Galilean version of pure Yang Mills theories with more general gauge groups. For generality of the discussion let's assume that the original Lorentzian gauge field 1-forms $A= A^a T_a$ take values in a semi-simple Lie-algebra $ \mathfrak{g}$ spanned by $T_a$, with $f_{abc}$ as structure constants. If vector space dimension of $ \mathfrak{g}$ is $\mathfrak{D}$, we will have here, a total of $\mathfrak{D}+1$ distinct Galilean limits of the gauge theory. Each of these limit sectors can be assigned one of the following $\mathfrak{D}$ dimensional vectors:
\bea{}
\Xi_{(p)} = (\underbrace{0,0,\dots , 0}_{\mathfrak{D}-p}, \underbrace{1,1,\dots , 1}_{p}) \qquad ~~ p=0, \dots, \mathfrak{D} 
\eea
We would denote the $a^{\mathrm{th}}$ component of $\Xi_{(p)}$ as $\Xi_{(p)}^a$, which can take values $0$ or $1$.

\subsection{Scaling of Fields} 
From now on we will concentrate on a given sector, let's say the Galilean $p_0^{\mathrm{th}}$ sector, $p_0$ however is arbitrary. In this sector, the scalar and the vector parts of the gauge fields descend from the relativistic gauge field through the following contraction:
\bea{}
A^a_t \longrightarrow \frac{\epsilon}{1+\epsilon-\Xi^a_{(p_0)}}A^a_t,		~~~~ 	A^a_i \longrightarrow \frac{\epsilon}{\epsilon+\Xi^a_{(p_0)}}A^a_i.
\eea
Clearly if $\Xi_{(p_0)}^a=1$, the contractions for the (Galilean) scalar and vector parts of the corresponding gauge field component in its most explicit form is (cf. \eqref{e-m_scaling}):
\bea{ymcontra1}
A^a_t \rightarrow  A^a_t, \, && \, A^a_i \rightarrow  \e A^a_i
\eea
On the other hand for $\Xi_{(p_0)}^a=0$, the following one holds:
\bea{ymcontra2}
A^a_t \rightarrow \e A^a_t, \, && \, A^a_i \rightarrow  A^a_i
\eea
With the observations \eqref{ymcontra1} and \eqref{ymcontra2}, we introduce a more convenient notation for the gauge fields, that would facilitate our forthcoming analysis. That is, the index $a$ in the range $1 \leq a \leq \mathfrak{D}-p_0$ will be denoted by capital Romans $I,J \dots$ and in the range $ \mathfrak{D}-p_0+1 \leq a \leq \mathfrak{D}$ will be Greeks $ \alpha, \beta \dots$. In this notation, \eqref{ymcontra1} and \eqref{ymcontra2} become respectively:
\bes
\bea{ymcontra}
&&  A^{\alpha}_t \rightarrow  A^{\alpha}_t, \, \qquad \, A^{\alpha}_i \rightarrow  \e A^{\alpha}_i\\
\mbox{and }&& A^I_t \rightarrow \e A^I_t, \, \qquad \, A^I_i \rightarrow  A^I_i.
\eea
\ees
The two extreme sectors, $p_0=\mathfrak{D}$ and $p_0= 0$ are the `Vanilla' limits, respectively corresponding to the pure `Electric' and the `Magnetic' limits for all gauge field components. 

\subsection{Equations of Motion} 
The next step will be to see the Galilean YM equations of motion in the $\mathfrak{D}+1$ distinct sectors. As per our convention, relativistic pure YM equation of motion is:
\bea{lorentz_ym_eom}
d \star F + g (A \wedge \star F - \star F \wedge A)=0
\eea
where the Hodge $\star$ is with respect to the Minkowski metric and $g$ is the coupling constant. Curvature or the field strength $F$, in the standard gauge theory formulation depends on $A$ through
$$F = dA + g A \wedge A.$$
Left hand side of \eqref{lorentz_ym_eom} is $ \mathfrak{g}$ valued tensor and hence represent $\mathfrak{D}$ number of equations, when stripped in the the Lie algebra basis. Moreover it can be transformed from a $D-1$ form to a single space-time index 1-form by a Hodge dual. As Galileanization breaks the covariance of $t$ and $x^i$, we will have two equations, for each free Lie algebra index. One would be space-time Galilean scalar and another vector. \eqref{sptmcontract} and \eqref{ymcontra} will be used for contracting the original relativistic equation. Subsequently multiplication by relevant power of $\e$ and taking the limit $ \e \rightarrow 0$ would give us the desired result.

As per our discussion above, the free Lie algebra index can fall in either of the two classes according to its contraction rule prescribed by the $ \Xi_{(p_0)}$ vector. These are described below.
\paragraph{Case 1: $\mathfrak{D}-p_0+1 \leq a \leq \mathfrak{D}$}
Let's look at the scalar equation first:
\bea{gymeom3}
\p^i \p_i A_t^{ \alpha}=0
\eea
While the space-time vector one is:
\bea{gymeom4}
&& \p_t \p_j A_t^{\alpha} + \p^i \left( \p_i A_j^{\alpha} - \p_j A_i^{\alpha} \right) \non\\ &&+g \bigg [f^{ \alpha}{}_{\beta \gamma} A_t^{\beta} \p_j A_t^{ \gamma} +  f ^{\alpha}{}_{ J K} \p^i \left(   A^J_i A^K_j \right) + f^{ \alpha}{}_{ J K } A^{iJ} \left( \p_i A^K_j - \p_j A^K_i\right) \bigg]=0
\eea

\subsection*{Case 2: $1 \leq a \leq \mathfrak{D}-p_0$}
Scalar equation:
\bea{gymeom1}
\p^i \p_i A_t^I -\p^i \p_t A_i^I + g f^{I}{}_{J\alpha} \left[\p^i \left(  A_i^J A_t^{ \alpha}\right) + A_i^J \p^i A_t^{\alpha}  \right]=0
\eea
Vector equation:
\bea{gymeom2}
\p ^i (\p_i A_j^I - \p_j A_i^I)=0
\eea

We see from the above equations that for a generic non-relativistic limit with some Electric and some Magnetic legs, the Galilean theory always contains interaction, corresponding to the usual momentum dependent vertex $\sim g$ in perturbation theory terminology. The quadratic gluon vertex $\sim g^2$ does not show up in the Galilean theory. Either of the scalar and the vector equation however trivializes.

It is also important to point out some non-generic cases to make connections to our earlier explicit construction of the $SU(2)$ theory. We had seen in the previous section that in some sectors of the $SU(2)$ theory (two out of the four), all interaction terms drop out of the equations of motion. A priori we did not have a reason to expect this. Now given the general structure of an arbitrary sector in the $SU(N)$ theory, we understand this better. From the equations above, we see that for the interaction terms to survive, the limits have to have 
\begin{itemize}
\item {\em{3 or more electric legs}}: then the first interaction term in \refb{gymeom4} survives. This is what happens in the pure electric case of the $SU(2)$ theory (Sec \ref{eee}). 
\item {\em{1 or 2 electric legs and 2 or more magnetic legs}}: then the interaction terms in \refb{gymeom1} survive as do the second and third terms of \refb{gymeom4}. This is what happens in the second skewed limit of the $SU(2)$ theory (Sec \ref{emm}). 
\item{\em{3 or more electric legs and 2 or more magnetic legs}}: then all interaction terms survive. This (obviously) does not have any $SU(2)$ example. 
\end{itemize} 
Note that in order to have interactions, the Galilean limit must have at least one electric leg. The purely magnetic limit always trivialises reducing to non-interacting copies of the Abelian magnetic sector. 

\subsection{Gauge Invariance of GYM: Limiting and Intrinsic analyses} 
Now that we have witnessed the invariance of the equations of motion for all Galilean sectors under full GCA, it would be interesting to address the issue of gauge invariance. As expected from the previous analysis for Electrodynamics and the $SU(2)$ case, we would notice that the amount of gauge freedom gets reduced in the Galilean case for more general non-Abelian theories too. We would start with the analysis for maximal gauge transformation from the relativistic case, take appropriate scaling to the Galilean regime, and then move over to a formulation from a more intrinsic point of view, as promised in the beginning of section \ref{su2section}. This is essentially in the framework outlined for the Electrodynamics case \refb{gt} which does not bear any reference to the relativistic theory.
\paragraph{Gauge invariance as a limit:} We would use the notation introduced in the previous subsection in order to deal with the several sectors of the Galilean theories. Galileanization will directly be effected on the gauge transformation rule of general 4-vector potential \refb{ymga}. As per the scheme outlined above in this section, the potential $A^{I}$ scales according to magnetic limit and $A^{\alpha}$ scales in electric limit. In addition to them, the $\alpha^{a}$'s should scale as  
\bea{sung} 
\alpha^{I}\rightarrow \epsilon \alpha^{I},\hspace{.2cm}
\alpha^{\alpha}\rightarrow \epsilon^{2} \alpha^{\alpha}
\eea
so as to keep the gauge transformation rules finite under the scaling.

Using these, the scaled rules for gauge transformation can be easily read off from \refb{sung}:
\begin{subequations}\label{sungm}
\bea{}\label{sungi} A^{I}_{t}\rightarrow A^{I}_{t}+\frac{1}{g}\partial_{t}\alpha^{I}+f^{I}_{\hspace{.2cm}\alpha J}A_{t}^{\alpha}\alpha^{J},\quad
 A^{I}_{i}\rightarrow A^{I}_{i}+ \frac{1}{g}\partial_{i}\alpha^{I}\\
\label{sungii} A^{\alpha}_{t}\rightarrow A^{\alpha}_{t},\quad
 A^{\alpha}_{i}\rightarrow A^{\alpha}_{i}+ \frac{1}{g}\partial_{i}\alpha^{\alpha}+f^{\alpha}_{\hspace{.2cm}IJ}A^{I}_{i}\alpha^{J} \eea
\end{subequations}
The equations of motion are invariant under this restricted set of gauge transformations.

Again a few remarks are in order to link up to our earlier $SU(2)$ construction. We see here that the non-Abelian nature of the gauge transformation survives in only there is at least one electric and two magnetic legs of the limit in question. This is why we had only one sector (EMM sector: Sec \ref{eem}) that displayed non-Abelian gauge transformations in the $SU(2)$ analysis. 

\paragraph{Intrinsic Gauge invariance:}
We would now detail the analysis for gauge invariance, done entirely from the intrinsic Galilean point of view through the procedure described in the conditions \ref{cond1}-\ref{cond3} of section \ref{gauge}. However we must point out that the ansatz for gauge transformation that we start off with is motivated by relativistic gauge fields. Let that be of the form:
\begin{subequations}
\begin{eqnarray}
\delta_{\Lambda} A^{a}_t = g^{-1}\, \partial_t \Lambda^{a}_1 + f^{a}{}_{bc}A^{b}_t \Lambda^{c}_1 \label{sungt1}  \\
\delta_{\Lambda} A^{a}_i = g^{-1}\, \partial_i \Lambda^{a}_2 + f^{a}{}_{bc}A^{b}_i \Lambda^{c}_2   \label{sungt2}
\end{eqnarray}
\end{subequations}
where $a$ can be either electric like $\alpha$ or magnetic like $I$ and hence a sum over $a$ implicitly a sum over $\alpha$ and $I$. Gauge parameters $\Lambda_{1,2}$ are chosen to be independent as in the Abelian case.

Let's start by inspecting if the transformation of the magnetic-like vector $\delta_{\Lambda}A_i^I = g^{-1}\, \partial_i \Lambda^{I}_2 + f^{I}{}_{bc}A^{b}_i \Lambda^{c}_2$ keep the equations of motion \refb{gymeom2} invariant. As is expected from the Abelian structure of this particular equation of motion, it is only invariant if we drop the non-Abelian part from the gauge transformation, ie allow  $\delta_{\Lambda}A_i^I = g^{-1}\, \partial_i \Lambda^{I}_2$. Next as we invoke the commutation of gauge transformation with Galilean boost, we get a condition very similar to the one of Electrodynamics: $\Lambda_1 = \Lambda_2$.

The same tests with the Electric-like scalar $A_t^{\alpha}$ reveals that it should not gauge transform at all. On the other hand, amount of gauge freedom allowed for the scalar in magnetic sector $A_t^I$ gets restricted from the general form of \refb{sungt1} to:
\begin{equation}
\delta_{\Lambda} A^{I}_t = g^{-1}\, \partial_t \Lambda^{I}_1 + f^{I}{}_{\alpha J}A^{\alpha}_t \Lambda^{J}_1 
\end{equation}
by the conditions. Now take the case of $A^\alpha_i$
\begin{equation}\label{last}
\delta_{\Lambda} A^{\alpha}_i = g^{-1}\, \partial_i \Lambda^{\alpha}_2 + f^{\alpha}{}_{bc}A^{b}_i \Lambda^{c}_2  
\end{equation}
Examination of (\ref{gymeom4}) reveals that the Abelian piece of (\ref{last}) will obviously keep the Abelian part of (\ref{gymeom4}) invariant (which is supported by our knowledge gathered from the Electrodynamics). On closer examination we see that, the interaction terms in (\ref{gymeom4}) have terms with $A^\alpha_t$ which as we've seen do not gauge transform and the other terms interacting terms are all in Magnetic sector and will be able to cancel the purely magnetic parts of (\ref{last}). So only the following gauge transformation keeps it invariant,
\begin{equation}
\delta_{\Lambda} A^{\alpha}_i = g^{-1}\, \partial_i \Lambda^{\alpha}_2 + f^{\alpha}{}_{IJ}A^{I}_i \Lambda^{J}_2  
\end{equation}
These results, as we see are in complete agreement with those found by taking the scaling limits \refb{sungm}. 

\subsection{Symmetries of EOM} 
For checking the invariance under GCA we would follow the same strategy as seen in $SU(2)$ case. Now to show the invariance of equations of motion of $SU(N)$ under GCA we have to relate the constants $(r,s)$ in term of $\Xi_{(p_{0})}$ first. 
\bea{changea}
r= - \Xi_{(p_{0})}, \quad s= \Xi_{(p_{0})} -1  
\eea
We would plug it back into \eqref{inf} in order to show the invariance under
$L^{(n)}$ and $M^{(n)}_{i}$. Since, we see that $\Xi_{(p_{0})} =0$ if the index $a$ is in the range $1 \leq a \leq \mathfrak{D}-p_0$ that is for (4.5a) and similarly it would be $\Xi_{(p_{0})} =1$ in the range $ \mathfrak{D}-p_0+1 \leq a \leq \mathfrak{D}$ for (4.5b).  
\\
\subsection*{Case 1: $1 \leq a \leq \mathfrak{D}-p_0$}
For the scalar equation, it is trivially invariant under  $M^{(n)}_{i}$. But under $L^{(n)}$, this equation is only invariant in space-time dimension 4:
\bea{}
\p^i \p_i [L^{n},A_t^I] -\p^i \p_t [L^{n},A_i^I] + g f^{I}{}_{J\alpha} \left( [L^{n},\p^i \left(  A_i^J A_t^{ \alpha}\right)] + [L^{n},A_i^J \p^i A_t^{\alpha}]  \right)\non\\
 = (\Delta -1)(n+1)[-nt^{n-1}\p_{i}A^{I}_{i}+gf^{I}_{J \alpha}t^{n}(A^{\alpha}_{t}\partial^{i}A^{J}_{i}+2A^{J}_{i}\p^{i} A^{\alpha}_{t})]\eea
For the space-time vector one: it is however invariant under f-GCA 
\bea{}
 \p^{i}\p_{i}[L^{(n)},A^{I}_{j}]-\p^{i}\p_{j}[L^{(n)},A^{I}_{i}]=0,\hspace{.2cm} 
\p^{i}\p_{i}[M^{(n)}_{l},A^{I}_{j}]-\p^{i}\p_{j}[M^{(n)}_{l},A^{I}_{i}]=0\eea 
\subsection*{Case 2: $\mathfrak{D}-p_0+1 \leq a \leq \mathfrak{D}$}
For the scalar equation, following the similar analysis as the previous case now gives
\bea{}
 \p^{i}\p_{i}[L^{(n)},A^{\alpha}_{t}]=0,\hspace{.2cm} \p^{i}\p_{i}[M^{(n)}_{l},A^{\alpha}_{t}]=0\eea   
For vector equation, $L^{(n)}$ keeps the equation invariant only in $\Delta =1$: 
\bea{}
 \p_t \p_j [L^{(n)},A_t^{\alpha}] + \p^i \p_i [L^{(n)},A_j^{\alpha}] - \p^i\p_j [L^{(n)},A_i^{\alpha}] +g f^{ \alpha}{}_{\beta \gamma} [L^{(n)},A_t^{\beta} \p_j A_t^{ \gamma}] + \non\\ g f ^{\alpha}{}_{ J K} [L^{(n)},\p^i \left(   A^J_i A^K_j \right)] 
 + g f^{ \alpha}{}_{ J K } [L^{(n)},A^{iJ} \left( \p_i A^K_j - \p_j A^K_i\right)]\non\\
= (\Delta -1)(n+1)[-nt^{n-1}\p_{j}A^{\alpha}_{t}+gf^{\alpha}_{\beta\gamma}t^{n}A^{\beta}_{t}\p_{j}
A^{\gamma}_{t}+gf^{\alpha}_{JK}t^{n}(A^{K}_{j}\p^{i}A^{J}_{i}\non\\-A^{J}_{i}\p_{j}A^{K}_{i}+2A^{J}_{i}\p^{i}A^{K}_{j})]
\eea
The EOM are trivially invariant under $M^{(n)}_{l}$. We have thus shown that the equations of motion of $SU(N)$ Yang-Mills theory exhibit infinite dimensional Galilean conformal invariance in $D=4$ in all of the different possible non-relativistic limits.   

\newpage

\section{Conclusions}

\subsection*{Summary}
In this paper, we have investigated Galilean limits of Yang Mills theories in some detail. We have seen that the colour index of the gauge field is responsible for a family of limits in the Galilean regime, generalising the Electric and Magnetic limits of the $U(1)$ theory earlier considered in \cite{Bagchi:2014ysa}. Galilean Yang-Mills theory thus consists of several sectors. We constructed the equations of motion in these different sectors and looked at the modified gauge invariances. We then proved that these EOM exhibit invariance under the full infinite dimensional GCA for all the different sectors in spacetime dimensions $D=4$.   Our initial explicit construction for the $SU(2)$ case led to atypical behaviour in several cases, like the absence of interaction terms in the equations of motion and the vanishing of non-Abelian structure in some of the four different limits. But we saw that when we looked at the details of the general $SU(N)$ story, we were able to resolve these apparent puzzles. We believe this observation of the infinite enhancement of symmetry in these interacting systems in the non-relativistic limit is very fascinating and possibly very useful as well.

\subsection*{Future Directions}
There are numerous directions of future work, some of which are currently under investigation. Here we provide a comprehensive list of these future directions. 

\paragraph{\it{Adding matter}:}The most obvious generalisation of the current work is to add matter fields. Since we are interested in non-relativistic conformal symmetries, it is natural to look at massless fields. In current work, which is in progress, we attempt to construct the non-relativistic analogue of scalar electrodynamics. We would be interested in also adding fermionic matter to construct a Galilean analogue of Quantum Electrodynamics, before going on to adding matter to the investigations of YM theories we have initiated in this work. It would be of interest to make connections to non-relativistic QCD theories \cite{Brambilla:2004jw, Soto:2006zs} which are effective field theories that have proved useful for heavy quarkonium physics. Quarkonia are bound states of quarks and are characterised by widely separated energy scales which makes effective field theory techniques very useful. The physics of the lower energy scales which are responsible for binding can be very difficult to access through perturbative calculations in QCD because the theory exhibits asymptotic freedom. Here effective field theory has been successfully employed to extract physics. We would like to link up these effective field theory discussions to the Galilean gauge theory descriptions we have put forward in this work. 

\paragraph{\it{Anomalies and actions}:} It would be very interesting thing to check whether the infinite dimensional Galilean conformal symmetry which has been the centrepiece of this paper only appears in classical Galilean gauge theories or if it miraculously also survives in the quantum regime. The natural expectation is that the symmetry would become anomalous, but the fact that there is actually an infinite dimensional symmetry to work with here makes us curious. We would like to initiate a study of anomalies in our theories. 

A natural obstacle is the absence of an obvious action formulation for our theories. This is something we would like to address in the near future. Let us make a few comments on this direction here. It is possible that an action formulation may be possible by introducing a set of auxiliary fields and following a procedure similar to that outlined in \cite{Verlinde:1993te}. Another possible way we could attempt an action formulation is by looking at Newton-Cartan structures. In a non-relativistic setting, where the metric on the whole of the spacetime degenerates, it is natural to adopt a geometric picture where the connections become the dynamical variables and talk between the base and fibres of the fibre-bundle structure that the spacetime now degenerates to. It may be possible to adopt fundamentally non-relativistic methods, like using the Newton-Cartan formulation, to understand the structure of Galilean gauge theories in terms of an action formulation. 

It is important to mention the very recent work \cite{Bergshoeff:2015sic} which came out while our paper was being readied for submission{\footnote{Also of interest is \cite{Bleeken:2015ykr} which employs Kaluza-Klein reductions to obtain a Newton-Cartan Maxwell dilaton system from a Newton-Cartan theory.}}. Here the authors consider systematic limits of minimally coupled relativistic theories to obtained Galilean field theories coupled to Newton-Cartan backgrounds. Their analysis for the massless Galilean fields is particularly of interest to us and we would be looking to incorporate their methods for the Galilean YM theory discussed in this paper. 

There has been some recent work on the construction of anomalies for Galilean field theories \cite{Jensen:2014hqa, Jain:2015jla}. But the procedure behind the formulation has been to start off with a relativistic theory with anomalies and do a Discrete Light Cone Quantisation (DLCQ). The dimensionally reduced theory is then Galilean invariant. So the process relates the relativistic theory to a non-relativistic theory in one lower dimension, unlike the process we have been following in our work here which relates Poincare and Galilean invariant theories in the same dimension. 

In \cite{Jensen:2014hqa}, the author also looks at conformal field theories with Galilean symmetry. However, the symmetry algebra considered has been the Schr{\"{o}}dinger symmetry as opposed to the GCA we have been looking at in this paper. The Schr{\"{o}}dinger Algebra (SA) and the GCA are fundamentally different.  The SA, unlike the GCA, is not obtained by an Inonu-Wigner contraction of the relativistic conformal algebra and has less generators than the GCA (there are no analogues of the spatial parts of the special conformal transformations for the SA). The SA has a mass extension, a central term which is the commutator between the boosts and the momenta, which is absent in the GCA. The GCA is thus the symmetry of massless or gapless non-relativistic systems which is closer to the relativistic conformal case. 

We wish to carry out an investigation of Galilean anomalies in the spirit of the current paper and our earlier work which relates relativistic and non-relativistic theories in the same dimension and then look at implications for the GCA as opposed to the SA. We believe that the GCA would have a central role to play when we consider renormalisation group flows in non-relativistic systems and would end up governing the fixed points of RG flows for Galilean field theories mirroring the role played by CFTs in Poincare invariant theories. Additionally, if we have the surviving infinite dimensional symmetry for these NR fixed points, we would be able to say more and the situation may be similar to 2d relativistic theories. 

\paragraph{\it{Supersymmetry and Integrability}:} The ultimate goal of our programme remains investigating the Galilean version of $\mathcal{N} =4$ $SU(N)$ supersymmetric YM. As we have mentioned earlier in the paper, the hope is that even if the Galilean conformal symmetries do not survive in the quantum version of Electrodynamics and in the current investigations of YM theories, like in the usual relativistic case, the conformal symmetries would also survive the quantum lift in the supersymmetric generalisation. In fact, we expect that we would find infinite Galilean super-conformal symmetries in the non-relativistic sector of $\mathcal{N} =4$ SYM. These infinite dimensional symmetries may indicate that there is a new integrable sub-sector of $\mathcal{N} =4$ SYM, different from the usual integrable planar sector. We would like to investigate integrability in detail when we look at this problem. It is perhaps worthwhile to study integrability already in the context of the non-supersymmetriec theories we have studied in this paper. 

\paragraph{\it{Other Classical Solutions}:} Solitons of YM theories, like instantons, monopoles, vortices and domain walls, to the best of our knowledge, have not been studied in the context of non-relativistic theories. It would be of great interest to look at these classical solutions both in terms of a non-relativistic limit and as solutions to the intrinsic Galilean theory. There is also the question of generalised electromagnetic duality in the Galilean YM theories. In the case of Galilean Electrodynamics, the electromagnetic duality exchanges the electric and magnetic sectors. Investigating Galilean versions of the Montonen-Olive dualities \cite{Montonen:1977sn} and the Witten effect \cite{Witten:1979ey} may lead to rich interplay in the various sub-sectors of the Galilean YM theory that we have discovered in this work. Needless to say, it would be more interesting to explore the strong-weak dualities in context of the supersymmetric version of the theory.  

\paragraph{\it{The Ultra-Relativistic limit}:} Before we close, we would like to advertise for upcoming work which is closely related to the current paper. We have studied the non-relativistic limit ($c \to \infty$) of YM theories in this paper. The ultra-relativistic limit ($c \to 0$) of gauge theories is also a very interesting sector to explore and this is one of the main features of \cite{conf-carroll}. These symmetries, curiously named Carrollian symmetries as opposed to Galilean ones, have made their appearance of late in the understanding of holography of flat spacetimes. Conformal Carroll groups are isomorphic to the Bondi-Metzner-Sachs groups \cite{Duval:2014uva} which are asymptotic symmetry groups in flat spacetimes at null infinity. Thus any putative dual field theories to flat spacetime living on the null boundary would be governed by conformal Carrolian symmetries. 

In two dimensions, the Conformal Carollian algebra (CCA) and the GCA are isomorphic and this ties up with the discussion at the beginning of the paper where we remarked that the 2d GCA was the symmetry structure that underlies the putative 2d field theory duals of 3d flat spacetimes \cite{Bagchi:2010eg}. The $c\to \infty \, (x^i \to \e x^i, t \to t, \e \to 0)$ limit and the $c\to 0 \, (x^i \to x^i, t \to \e t, \e \to 0)$ limit are related by flipping the spatial and time directions. In 2d, since there is only one spatial direction, the $x \leftrightarrow t$ does not change the algebra obtained in the limit which has one contracted and one uncontracted direction. In higher dimensions, the isomorphism between the CCA and the GCA is broken because now since the CCA contains only one contracted direction as opposed to the $(d-1)$ contracted directions of the $d$-dimensional GCA. Another feature that the two algebras do not seem to share is the infinite dimensional lift in arbitrary dimensions. In \cite{Bagchi:2012cy}, we were able to construct an infinite lift for the 3d CCA (or the BMS$_4$) by methods similar to what was done for the GCA in \cite{Bagchi:2009my}. But the construction does not seem to have a natural generalisation to field theories in spacetimes higher than $d=3$. 

In $d=4$, the ultra-relativistic limit of electrodynamics and YM theories lead to Conformal Carrollian field theories (CCFTs) which are putative duals to 5d Minkowski spacetimes. In a companion paper which would appear shortly \cite{conf-carroll}, we delve into the details of this construction following methods outlined in our current work. The paper would also contain some details about the representation theory of CCFTs in various dimensions, especially some surprises in 3d CCFTs which would be linked to quantum gravity in 4d Minkowski spacetimes. 

\bigskip

\subsection*{Acknowledgements}
It is a pleasure to acknowledge discussions with Shankhadeep Chakrabortty, Daniel Grumiller, Hong Liu, Joan Simon.  
We thank the organisations that supported this research: the Fulbright foundation (AB), the Inspire Faculty award of the Department of Science and Technology, India (RB), 
NAMASTE India-EU fellowship (AM).   

\bigskip \bigskip

\appendix
\section{GCA Negative modes and Invariance}
In this appendix, we look at the issue of invariance of the equations of motion under the negative modes of the GCA. In order to get an intuition of the workings of the GCA, we start off in $D=2$ and in the relativistic theory. We then use the method of contraction to get to the non-relativistic answer in $D=2$. We will then motivate the answer for general dimensions. 
\subsection*{Action of negative Virasoro modes}
We want to look at a method to find the action of negative Virasoro modes on the holomorphic part of primary of dimension $h$ in 1+1 d.
For $n\geq-1$, this is given by, 
\be{1}
[\mathcal{L}_{n},\phi_h(z)]= \left\{z^{n+1}\partial_z + h(n+1)z^{n}\right\}\phi_h(z)
\ee
In terms of the mode expansion 
\be{}\phi_h(z)= \sum_{n \in \mathbb{Z}}z^{-n-h}\phi_n,
\ee
the above equation \refb{1} becomes
\begin{equation}
[\mathcal{L}_{n},\phi_m]= \{n(h-1)-m\}\phi_{n+m}
\end{equation}
To find out the action of $\mathcal{L}_{-n}$ we conjugate both sides using $\mathcal{L}_n^\dagger=\mathcal{L}_{-n}$ and $\phi_{n}^\dagger=\phi_{-n}$,
\begin{equation}
[\mathcal{L}_{-n},\phi_m]= \{-n(h-1)-m\}\phi_{-n+m}
\end{equation}
This immediately implies, for $n\geq-1$
\begin{equation} \label{2}
[\mathcal{L}_{-n},\phi_h(z)]= \left\{z^{-n+1}\partial_z + h(-n+1)z^{-n} \right\} \phi_h(z)
\end{equation}
The anti-holomorphic part follows exactly in the same way. 
\subsection*{Action of negative 2d GCA modes}
In $D=2$, the GCA can be obtained as a contraction of two copies of the Virasoro algebra:
\be{}
\L_n + \bL_n = L_n, \quad \L_n - \bL_n = \frac{1}{\e} M_n
\ee
In $D=2$, the highest representations of the GCA are also obtained as a limit of the Virasoro highest weight representations and these GCA representations are labelled by their weights under $L_0$ and $M_0$. 
\be{}
L_0 |h_L, h_M \> = h_L |h_L, h_M \>, \quad M_0 |h_L, h_M \> = h_M |h_L, h_M \>, 
\ee
where $h_L$ and $h_M$ are related to the Virasoro highest weights by 
\be{}
h_L = h+ \bar{h}, \quad h_M = \e ( h - \bar{h}).
\ee
Primary states are defined in the usual way in this representation, viz. the action of the positive modes of the GCA annihilate the state. Given the action of the negative modes in the relativistic case, we can immediately write the action of the negative GCA modes. This lets us write the action of the GCA generators by using the expression (\ref{1}) and its anti holomorphic counterpart on a primary operator $\phi(t,x)${\footnote{Note that operators and states are linked by a state-operator correpondence $|\phi\> = \phi(0) |0\>$}}.  This gives us for $n\geq-1$,
\bea{}
\delta_{L_n} \phi(t,x) &=&[L_n,\phi(t,x)] \non \\
&=& \left[(t^{n+1}\partial_{t}+ (n+1) t^nx\partial_x + (n+1)(h_L t^n - n h_M t^{n-1}x)\right]\phi(t,x) \\
\delta_{M_n} \phi(t,x) &=& [M_n,\phi(t,x)] \non \\ &=& \left[-t^{n+1}\partial_{x}+ (n+1)t^n h_M \right] \phi(t,x)
\eea
The action of the GCA negative modes is clear once we know the action of the negative Virasoro modes. We just have to use (\ref{2}) and it's anti holomorphic part instead of (\ref{1}). This gives for $n\geq-1$,
\bea{}
\delta_{L_{-n}} \phi(t,x) &=& [L_{-n},\phi(t,x)]  \\
&=& \left[t^{-n+1}\partial_{t}+ (-n+1) t^{-n}x\partial_x + (-n+1)(h_L t^{-n} +n h_M t^{-n-1}x)\right] \phi(t,x) \non\\
\delta_{M_{-n}} \phi(t,x) &=& [M_{-n},\phi(t,x)] \non \\ &=&  \left[-t^{-n+1}\partial_{x}+ (-n+1)t^{-n} h_M \right] \phi(t,x)
\eea
So we see that like in the case of the relativistic 2d CFT, for a 2d GCFT the action of the negative GCA modes on the primary fields are given by a replacement of $n \to -n$ on the right hand side of the equations for the positive modes. 

\subsection*{Negative modes of GCA in general dimensions}
In our analysis of electrodynamics in \cite{Bagchi:2014ysa} and YM theories in this paper, we have resorted to looking at the representations of the GCA labelled by the dilatation and the angular momentum generator. The states of interest are build on scale-spin primaries instead of scale-boost primaries which we have just discussed for the two dimensional example above. For the scalar theory in dimensions $D>2$, the action of the operators on the states would reduce to that of the $h_M=0$ sector of the theory discussed above:
\bea{}
\lb L^{(-n)},\phi(t,x^i) \rb &=& \left[t^{-n+1}\partial_{t}+ (-n+1) t^{-n}x^i\partial_i + (-n+1)\Delta t^{-n}\right] \phi(t,x^i) \\
\lb M^{(-n)}_i,\phi(t,x^i)\rb &=& -t^{-n+1}\partial_i  \phi(t,x^i)
\eea
Note that here we have replaced $h_L = \Delta$. Building on this intuition, we postulate that when we are looking at the non-trivial scale-spin primaries, the $n \to - n$ change would also hold for general dimension. Admittedly this is a conjecture which merits a proof,- but we have seen that it is motivated by the 2d example as well as the scalar example in general dimensions. Hence we feel justified making this claim. Under this assumption, it can be easily checked that the invariances of all the equations of motion for the negative modes will go through without any problems.

\end{document}